\documentclass[preprint]{aastex}
\usepackage{color}
\usepackage{graphicx}	
\usepackage{amsmath}	
\usepackage{amssymb}	
\usepackage{natbib}
\usepackage{enumerate}

\received{XXX}
\revised{YYY}
\accepted{ZZZ}

\shorttitle{Chemical variations across the TMC-1 boundary}
\shortauthors{Chen et al.}

\begin{document}

\title{{Chemical variations across the TMC-1 boundary}: molecular tracers from translucent phase to dense phase}

\author[0000-0002-9703-3110]{Long-Fei Chen}
\email{chenlongf@nao.cas.cn}
\affil{National Astronomical Observatories, Chinese Academy of Sciences, 20A Datun Road, Chaoyang District, Beijing 100101, China}

\author[0000-0003-3010-7661]{Di Li}
\email{dili@nao.cas.cn}
\affil{National Astronomical Observatories, Chinese Academy of Sciences, 20A Datun Road, Chaoyang District, Beijing 100101, China}
\affil{University of Chinese Academy of Sciences, Beijing 100049, China}
\affil{NAOC-UKZN Computational Astrophysics Centre, University of KwaZulu-Natal, Durban 4000, South Africa}

\author[0000-0003-4811-2581]{Donghui Quan}
\affil{Department of Chemistry, Eastern Kentucky University, 521 Lancaster Ave., Richmond, KY 40475, USA}

\author[0000-0003-4843-8944]{Xia Zhang}
\affil{Xinjiang Astronomical Observatory, Chinese Academy of Sciences, No.~150 Science 1-Street, Urumqi 830011, China}

\author[0000-0002-0040-6022]{Qiang Chang}
\affil{School of Physics and Optoelectronic Engineering, Shandong University of Technology, Zibo 255000, China}

\author[0000-0003-2090-5416]{Xiaohu Li}
\affil{Xinjiang Astronomical Observatory, Chinese Academy of Sciences, No.~150 Science 1-Street, Urumqi 830011, China}

\author[0000-0002-9511-7062]{Lin Xiao}
\affil{College of Science, Changchun Institute of Technology, Changchun, Jilin 130012, China}

\begin{abstract}
We investigated the chemical evolutions of gas phase and grain surface species 
across the Taurus molecular cloud-1 (TMC-1) filament from translucent phase to dense phase.
By comparing observations with modeling results from an up-to-date chemical network, 
we examined the conversion processes for the carbon-, oxygen-, nitrogen- and sulfur-bearing species, 
i.e.~from their initial atomic form to their main molecular reservoir form 
both in the gas phase and on the grain surface. 
The conversion processes were found to depend on the species and A$_V$.
The effect of initial carbon to oxygen elemental abundances ratio (C/O) by varying O on the chemistry was explored, 
and an initial carbon elemental abundance of 2.5 $\times$ 10$^{-4}$ and a C/O ratio of 0.5 
could best reproduce the abundances of most observed molecules at TMC-1 CP, where more than 90 molecules have been identified.
Based on the TMC-1 condition, we predicted a varied grain ice composition during the evolutions of molecular clouds, 
with H$_2$O ice as the dominant ice composition at A$_V$ $>$ 4 mag, CO$_2$ ice as the dominant ice composition 
at A$_V$ $<$ 4 mag, while CO ice severely decreased at A$_V$ around 4--5 mag.

\end{abstract}

\keywords{astrochemistry (75) -- Chemical abundances (224) -- Interstellar medium (847) -- Interstellar molecules (849)}

\section{Introduction}
Understanding the chemical evolution during the transformation of atomic gas to molecular gas 
of the clouds is crucial to construct a complete picture of the evolutions of the interstellar medium (ISM).
In the transition phase of molecular clouds, 
such as from the diffuse phase to the translucent phase and then to the dense phase, 
physical changes of the environments affect the local chemical states.
Molecules and their associated chemistry, such as hydrocarbon chemistry, sulfur-containing molecular chemistry and
complex organic molecular chemistry, are imprints of the evolutions of ISM.
The choice of suitable tracers that represent the chemistry could
reflect the changes of physical processes as well, thus the evolution of star formation \citep{Aikawa2001}.

One process in such a transitional phase is the transformation of atomic hydrogen to
molecular hydrogen on the grain surface at low temperatures \citep{Cuppen2006}. 
Another set of widely recognized process is the transformations between C$^+$, C and CO 
during the transition from the diffuse phase to the dense phase of a cloud \citep{Lee1996}. 
The transformations of species from atomic to molecular form, or from simple to complex form indicate that the chemical
complexity is increasing with the evolution of the clouds. 
Thus, it is worth exploring other kinds of molecules, such as sulfur-bearing molecules
and grain surface species, as tracers in these transitional phases.
Moreover, the boundaries defined by different tracers may be characterized by
different physical conditions, such as density or temperature \citep{Goldsmith2010}.
Thus, the study of the chemistry of different molecules could benefit our understanding of 
the physical environments.

The elemental sulfur abundance is severely depleted in dark clouds while maintaining its atomic form in 
diffuse clouds \citep{Vidal2017, Laas2019}. Thus, there should be a transformation for sulfur-bearing
species if we analogize the sulfur chemistry with the carbon chemistry.
As stated above, dust grains are important for the formation of H$_2$. 
Other species could also stick onto the surface of grains, diffuse among the sites and recombine to form new species. 
Water ice is generally observed on the line-of-sight dense clouds toward background stars. 
At the diffuse ISM condition, water ice is expected by the laboratory experiments \citep{Potapov2020}. 
However, there was no strong evidence for the existence of solid-state water in the diffuse-dense transition \citep{Poteet2015}.
The presence of water ice is important for the depletion of elemental oxygen in the gas of diffuse ISM, which
could be explained if the missing oxygen is in the oxygen-bearing solid form in the grain ice mantles \citep{Poteet2015}.
On the other hand, the detection of larger complex molecules, such as CH$_3$OH in the diffuse
and translucent clouds \citep{Thiel2017, Liszt2018} could also be explained by the grain surface origin.
The chemical composition of grain ice mantles in the diffuse-dense transition is poorly constrained 
because of the observational difficulty in the identification of interstellar ice mixture. 
\citet{Esplugues2019} studied the ice chemistry in the photodissociation regions (PDRs) 
where high UV radiation is present, which results in poor ice compositions (i.e~less than one monolayer when A$_V$ $<$ 2 mag).
With respect to the PDRs, the UV radiation in the outer extended regions of the dark clouds is much lower, 
so that a relatively rich ice composition could be developed in that region. 
Thus, the comparison of the ice composition between the outer and inner regions of the dark clouds 
could be more meaningful compared with the PDRs.

The Taurus molecular cloud (TMC) has been chosen as the target by \citet{Goldsmith2010}, \citet{Xu2016a} and \citet{Xu2016b} 
to study the boundary conditions indicated by H$_2$ emission, OH and CH physicochemical properties, respectively.
Taurus molecular cloud-1 (TMC-1) is one of a nearest clouds with a distance of $\sim$140 pc \citep{Elias1978}.
It is in a place where the low-mass star formation process is ongoing
and is currently close to gravitational equilibrium \citep{Feher2016}.
It is also in an environment that is less affected by the external radiation field \citep{Xu2016a} compared with the PDRs,
making it an ideal source for the study of chemical evolutions.

Observationally, many species have been detected in the diffuse and translucent phases, 
indicating that chemical complexity has been developed in these regions \citep{Snow2006, Liszt2018}.
With relatively low density and low extinction, many small hydrocarbons
and ions survived \citep{Lucas2000, Liszt2018, Thiel2019}.
With the accretion of the gas, the molecular species diversity also increases.
Examples are the detection of cyanopolyyne species \citep{Suzuki1992}, methanol and even more
complex organic molecules (COMs) \citep{Soma2015, Soma2018} in the dense part of TMC-1.
More recently, more than two dozens of molecules have been detected in TMC-1, 
including the long carbon chain molecules 
\citep{Xue2020, Loomis2021, Shingledecker2021, Cernicharo2021f, Cabezas2021b, Marcelino2021}, 
circular and aromatic molecules \citep{McGuire2021, Burkhardt2021, Cernicharo2021c},
unsaturated hydrocarbons \citep{Agundez2021a, Cabezas2021a, Cernicharo2021e},
oxygen-bearing COMs \citep{Agundez2021b, Cernicharo2020b, Cernicharo2021a}, 
sulfur-bearing molecules \citep{Cernicharo2021b, Cernicharo2021d},
and nitrile anions \citep{Cernicharo2020a}.
A systematic check of the chemistry based on this large number of observed molecules could give a
new insight into the constraints of the chemical reaction network.

Many observational and theoretical studies have focused either on
the chemistry in diffuse or translucent part \citep{Lucas2000, Liszt2018, Thiel2019} with extinctions
less than $\sim$5 mag or on the dense part of TMC-1 CP (the cyanopolyynes emission peak) or TMC-1 NH$_3$
(the ammonia emission peak) \citep{Soma2015, Soma2018, Loison2017, Majumdar2017, Maffucci2018}.
It is interesting to explore the chemical variations from the diffuse
to the dense part of the TMC-1 \citep{Lee1996, Fuente2019}.
\citet{Turner2000} studied the chemical variations from the diffuse to translucent clouds
at extinction in the range from $\sim$1 to 5 mag by using the pure gas-phase chemical models.
More recently, \citet{Federman2021} studied the changes of the physical conditions 
from diffuse gas to molecular gas for four different line-of-sights toward background star 
in TMC by multi-wavelength observations. \citet{Bron2021} explored the different tracers
of the ionization fraction in translucent and dense gas.
In this paper, by combining the most recent observationally derived physical properties of TMC-1 and 
the state-of-the-art astrochemical simulations from diffuse to dense phase
that across the TMC-1 boundary with the extinctions range from 1 to 20 mag, 
we investigate the chemical variations in these transitional regions
and study the chemical properties of most important molecules that may serve as the tracers of these regions.

The organization of this paper is as follows. In Section \ref{sec.model}, we describe the
changes of the physical conditions from the ridge of TMC-1 filament to the outer extend regions, 
and the chemical models used for the evolution of the chemistry across a span of virtual extinction.
Section \ref{sec.result} shows the results at the dense part of the TMC-1 CP by comparison with
observations to constrain the chemical model parameters, and then shows the chemical variations as a function
of extinction across the TMC-1 boundary. 
Finally, we discuss our predicted results in Section \ref{sec.discuss} 
and highlight major findings in Section \ref{sec.conc}.

\section{Physicochemical models of TMC-1 boundary}\label{sec.model}

\subsection{Physical parameters description}
Along and near the TMC-1 filament, there are three observational emission peaks,
namely TMC-1 CP (the cyanopolyynes emission peak), TMC-1 NH$_3$ (the ammonia
emission peak) and TMC-1 C (the dust continuum peak).
As shown in Figure 1 of \citet{Fuente2019}, starting from the three emission peaks, 
there are six cutting points extending horizontally to the edge of the molecular cloud.
They represent the transitional regions from the inner dense phase to the outer translucent phase.
The physical conditions of the model are extracted from \citet{Fuente2019} in which for each point 
the dust temperature, gas temperature, H$_2$ column density and volume density were derived. 
The dust temperature is derived from the \textit{Herschel} Gould Belt Survey, and the gas temperature
and n(H$_2$) are derived by fitting the molecular lines of CS and its isotopologues.
The extinction is derived by $N(H_2) = A_V \times 10^{21}$.
The total hydrogen nuclei density n$_H$ equals to n(H) + 2n(H$_2$).
Here we assume the majority of matter is occupied by H$_2$, thus n$_H$ $\approx$ 2n(H$_2$).
We list these physical parameters in Table \ref{tab.para}.

\begin{table}
\caption{Physical conditions along the three horizontal lines starting from the three emission peaks in TMC-1 filament.}
\label{tab.para}
\begin{tabular}{lllll}
\hline \hline
             & T$_{dust}$ & T$_{gas}$ & A$_V$ & n$_H$  \\
             & (K)      & (K)     & (mag) & ($\times$ 10$^4$ cm$^{-3}$) \\
\hline
TMC-1 CP     & 11.92 & 9.7   & 18.20 & 3.0 \\
offset 1     & 12.00 & 10.2  & 16.71 & 4.6 \\
offset 2     & 12.24 & 11.2  & 13.74 & 7.4 \\
offset 3     & 13.16 & 12.5  & 7.27  & 0.3 \\
offset 4     & 13.86 & 16.0  & 4.77  & 0.54 \\
offset 5     & 14.39 & 14.7  & 3.25  & 0.32 \\
\hline
TMC-1 NH$_3$ & 11.70 & 11.8  & 16.97 & 4.0 \\
offset 1     & 11.79 & 10.2  & 15.58 & 2.4 \\
offset 2     & 12.12 & 12.7  & 12.88 & 3.0 \\
offset 3     & 13.10 & 11.9  & 10.04 & 0.58 \\
offset 4     & 13.78 & 11.4  & 4.04  & 0.32 \\
offset 5     & 14.00 & 13.5  & 3.00  & 0.58 \\
\hline
TMC-1 C      & 11.26 & 8.5   & 19.85 & 9.0 \\
offset 1     & 11.32 & 10.3  & 18.47 & 8.6 \\
offset 2     & 11.67 & 11.6  & 13.34 & 2.38 \\
offset 3     & 13.13 & 11.1  & 4.79  & 0.58 \\
offset 4     & 14.08 & 13.5  & 2.20  & 1.1 \\
offset 5     & 14.53 & 13.5  & 1.63  & 0.52 \\
\hline
\end{tabular}
\\

Notes.\\
Physical parameters from the three emission peaks to the outer extend regions of TMC-1.
For each set of parameters in TMC-1 CP, NH$_3$ and C, 
the first row corresponds to the observational molecular emission peak in the TMC-1 filament,
while the other rows are the horizontal extensions starting from the emission peak.
Data are taken from \citet{Fuente2019}
(the T$_{dust}$ and A$_V$ in the sixth row of TMC-1 NH$_3$ set have been corrected by the author).
\end{table}

\subsection{Chemical model description}
We use the two-phase (gas phase and grain surface) model of NAUTILUS \citep{Ruaud2016}
to perform the astrochemical simulations.
The full gas-grain reaction network is the same as used in \citet{Vidal2017}, which is considered 
the up-to-date sulfur chemistry to solve the elemental sulfur depletion problem, and has been
tested at the TMC-1 CP.
This network is based on kida.uva.2014 \citep{Wakelam2015} and \citet{Garrod2006}.
We also modified the network with additional reactions related to C$_3$H and C$_3$H$_2$ isomers from \citet{Loison2017},
namely linear and cyclic configuration, which are widely detected in the interstellar
medium including in the diffuse and translucent environments \citep{Turner2000, Gerin2011}.

Basically, kida.uva.2014 is the gas phase network, which includes various types of reactions,
such as ion-neutral and neutral-neutral reactions, cosmic-ray and UV-photon dissociations.
In addition to the gas phase chemical reactions, there are also grain surface reactions,
and interchange processes between gas phase and grain surface species.
The gas phase species can be adsorbed to and diffuse on the grain surface and react with other surface species.
In addition to including the thermal desorption and cosmic-ray desorption of grain surface species,
the network also includes the chemical desorption, which is the desorption by exothermicity
of surface reactions \citep{Garrod2007} with an efficiency of 0.01.
The ratio of diffuse to desorption energy of a grain surface species
is set as the default value of 0.4 \citep{Garrod2006, Chang2012}.
As for the strength of the interstellar radiation field,
the cosmic-ray ionization rate is set as 5 times the standard value of 1.3 $\times$ 10$^{-17}$ s$^{-1}$,
and the UV factor is set as 3.8 in habing unit. 
These two values are consistent with the PDR model results \citep{Fuente2019} where different values of cosmic-ray
ionization and UV factor are used to fit the gas ionization fraction and the relation between dust temperature
and A$_V$, respectively.
We do not distinguish the difference between the grain surface and grain ice mantle species.
As for other parameters used in the model, the gas-to-dust mass ratio is 100,
and we assume a spherical silicate dust grain, with a commonly used radius of
0.1 $\mu$m and a density of 3 g cm$^{-3}$.
We also assume the number density of grain surface sites is 1.5 $\times$ 10$^{15}$ cm$^{-2}$.

Due to the low extinction in the boundary of the TMC-1 filament as shown in Table \ref{tab.para}, 
the shielding for H$_2$ and CO should be important.
NAUTILUS includes the self-shielding of H$_2$ \citep{Lee1996},
CO self-shielding and cross-shielding by H$_2$ and dust \citep{Visser2009},
and the shielding of N$_2$ by H$_2$ and N$_2$ \citep{LiXiaohu2013}.

The initial elemental abundances are chosen from \citet{Semenov2010} with the exception of C$^+$, O and S$^+$
because of the importance of these elements on the chemistry.
\citet{Agundez2013} has shown that a carbon-poor, oxygen-rich condition (C/O = 0.55)
could reproduce over 60 percent modeled species within one order of magnitude 
when compared with observed molecules at TMC-1 CP. 
While for the cyanopolyynes species (HC$_n$N, n = 3, 5, 7, 9),
they are more favored in the opposite condition (with C/O = 1.2) as shown by \citet{Maffucci2018}.
On the other hand, the sulfur element depletion is an unsolved problem \citep{Jenkins2009, Vidal2017, Vastel2018}.
It is known that sulfur remains its ionized atomic form in diffuse and
translucent interstellar medium \citep{Jenkins2009, Laas2019} at an
abundance of its cosmic value, while it is severely depleted in the gas phase of the
dense clouds or cores in almost three orders of magnitude \citep{Agundez2013, Laas2019}.
This problem is raised partially because of the insufficient sulfur chemical reactions in the network,
so that observational values can only be reproduced when the depleted initial elemental sulfur abundance in dense clouds is chosen.
With an update of the sulfur reactions, \citet{Vidal2017} showed that an undepleted elemental sulfur abundance, 
which is corresponding to its cosmic value, could also fit the observations in the dense clouds condition.
Thus, a universal value of the initial sulfur elemental abundance is used throughout the simulations range from
different transitional phase of the clouds.
Table \ref{tab.elem} shows the initial elemental abundances.

\begin{table}
\caption{Initial elemental abundances relative to hydrogen nuclei.}
\label{tab.elem}
\begin{tabular}{lc}
\hline \hline
Element     & Abundance  \\
\hline
H$_2$       & 0.5 \\
He          & 9.00(-2) \\
C$^+$$^a$   & 1.2(-4), 1.8(-4), 2.5(-4) \\
O$^b$       & varied \\
N           & 7.60(-5) \\
S$^+$$^c$   & 1.50(-5) \\
Si$^+$      & 8.00(-9) \\
Mg$^+$      & 7.00(-9) \\
Fe$^+$      & 3.00(-9) \\
Na$^+$      & 2.00(-9) \\
Cl$^+$      & 1.00(-9) \\
P$^+$       & 2.00(-10) \\
\hline
\end{tabular}
\\

Notes.\\
a(-b) represents a $\times$ 10$^{-b}$.\\
$^a$ Three different values are tested, see Section \ref{sec.TMC-1CP}.
The low, intermediate and high values are taken from 
\citet{Semenov2010}, \citet{Agundez2013} and \citet{Przybilla2018} respectively. \\
$^b$ Calculated with varied C/O ratio range from 0.4 to 1.4. \\
$^c$ Cited from \citet{Vidal2017}. \\
All other values are cited from \citet{Semenov2010}.
\end{table}

\section{Results and analysis}\label{sec.result}
We show the results in this section. We first constrain the initial carbon and oxygen elemental abundances
for the chemical models by comparing the observational molecular abundances at the TMC-1 CP 
with our modeled abundances.
The TMC-1 CP is one of the famous targets for many observations.
There are over 90 molecules have been detected at this location.
See \citet{Ohishi1992IAUS, Kaifu2004, Adande2010, Agundez2013, Gratier2016, McGuire2017} 
and recent observational references in the previous section.
We divide these molecules into four groups according to their elemental characteristics 
and the role they played in the reaction network.
The S group represents all species that include sulfur element.
The N group represents nitrogen-containing species excluding S group species.
The O group represents oxygen-containing species excluding S and N group species.
Finally, the C group represents the remaining species.
They are mainly the hydrocarbons that only include the hydrogen and carbon element.
This group also includes the atomic carbon and diatomic carbon.

\subsection{Chemical variations at the TMC-1 CP emission peak}\label{sec.TMC-1CP}
As introduced previously, the initial carbon to oxygen elemental abundances ratio (C/O)
is important for chemistry in ISM. In this section, we first explore the variations of this ratio on the chemistry. 
We keep the initial carbon elemental abundance a fixed value and vary the C/O to derive the initial
oxygen elemental abundance. To explore the variations of the initial carbon abundance, we use three different
values corresponding to low, intermediate and high abundance, 
which are 1.2 $\times$ 10$^{-4}$ \citep{Semenov2010}, 1.8 $\times$ 10$^{-4}$ \citep{Agundez2013} 
and 2.5 $\times$ 10$^{-4}$ \citep{Przybilla2018} respectively,
while the C/O ratio is varied from 0.4 \citep{Lee1996} to 1.4 \citep{Agundez2013} with a step size of 0.1.
In the following, we show our best fit model results, 
followed by a discussion on the chemistry due to the variations of C/O ratio.

\subsubsection{Best fit model}\label{sec.fitmodel}
By comparing the observational abundances of the molecules with the calculations from the models,
we could determine the best fit model and the chemical evolutionary timescale of the molecular cloud.
We use the following Equation \ref{eq.rms}, which is known as the distance of disagreement \citep{Vidal2017},
to calculate the mean logarithmic difference for all detected molecules at every time step.

\begin{equation}
D(t) = \frac{\sum_{i} | log(X_{mod,i}(t)) - log(X_{obs,i}) |}{N_{obs}}
\label{eq.rms}
\end{equation}
where $X_{mod,i}(t)$ and $X_{obs,i}$ is the modeling abundance for species $i$ at time $t$
and observed abundance, respectively, and $N_{obs}$ is the total number of the observed species.

\begin{figure}
\includegraphics{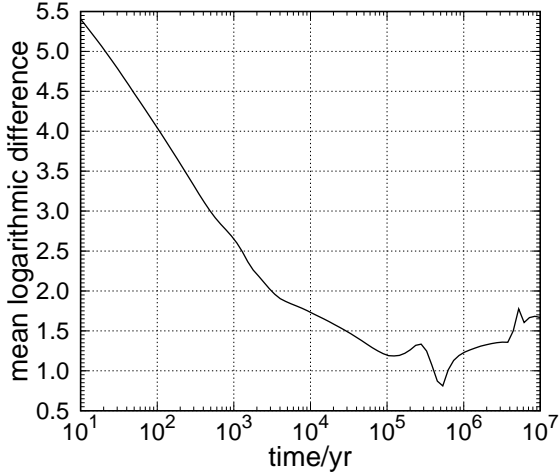}
\caption{The mean logarithmic difference for all detected molecules as a function of time
at the TMC-1 CP for the best model with an initial carbon abundance of 2.5 $\times$ 10$^{-4}$ and
a C/O ratio of 0.5.
}
\label{fig.rms}
\end{figure}

Figure \ref{fig.rms} shows the calculated results for the initial carbon abundance 
of 2.5 $\times$ 10$^{-4}$ and a C/O ratio of 0.5. It can be seen that the best fit time
is located at $\sim$5.3 $\times$ 10$^5$ year. This best fit time is in well agreement within
the range of 4.2$\pm$2.4 $\times$ 10$^5$ year estimated from the CO depletion timescale toward
13 cores in Taurus \citep{Pineda2010}.
Moreover, in order to show the diversity of the abundance between observational value and
modeling value of each species, we plot in Figure \ref{fig.TMC1-CP} the logarithmic abundance ratio difference
for each observed species.
In the figure, the left endpoint and right endpoint for each molecule is the minimum and maximum value
in the time range of 10$^4$ to 10$^7$ year respectively. The red cross mark corresponds to the time
at best agreement, i.e.~5.3 $\times$ 10$^5$ year. The vertical grey area between -1 and +1
represent one order of magnitude lower and higher respectively when compared with the observed values.

In Figure \ref{fig.TMC1-CP}, there are 66 out of 93 species, 
which is 71 percent whose logarithmic abundance is within the range of -1 to +1 of observation at the best fit time.
This percentage is similar to the results of earlier studies 
with a total observed species of $\sim$60 \citep{Agundez2013, Wakelam2015}.
Specifically, for the 23 O group molecules, C$_3$O and CH$_3$O are about two orders of magnitude
overproduced, while the large COMs (CH$_3$CHO, HCOOCH$_3$ and CH$_3$OCH$_3$) are underproduced.
For the 22 C group molecules, there are 8 species out of the range from -1 to +1.
Most of these out-of-range species are C$_n$H$^-$ and C$_n$H$_2$ with n = 4, 6, 8. Note that 
all of the small hydrocarbons with fewer than 3 carbon elements are much closer to the observations
than the large hydrocarbons. This trend is also seen in other models, which may indicate that 
the inefficient destruction pathways for the larger hydrocarbons when they grow complex.
Nitrogen-bearing molecules in the N group are the most detected species in the TMC-1 CP. 
Among them, there are 5 out of 28 that are outside the range from -1 to +1.
They are CN, C$_5$N, H$_2$CN, NH$_2$CHO and HCN, 
which are arranged from large to small disagreement with respect to their corresponding observed values.
Comparing with the sequential arrangement for the overproduced small hydrocarbons and large hydrocarbons,
there is no clear such an arrangement for these nitrogen-bearing molecules.

Finally, about a half of S group species (9 out of 20) are outside the range of one order of magnitude
when compared with their observational values at the best fit time.
Among them, CS$^+$ is the most underproduced species, the predicted peak fractional abundance with respect to H
is 2.2 $\times$ 10$^{-11}$ at $\sim$250 year and severely decreased after that. 
While the observed abundance of CS$^+$ is 1.5 $\times$ 10$^{-10}$.
As a comparison with CS$^+$, CS is overproduced by two orders of magnitude at its best fit time. 
The predicted peak abundance of CS is 1.9 $\times$ 10$^{-6}$ at $\sim$3.9 $\times$ 10$^4$ year and also 
decreases after that. While the observed abundance for CS is 1.5 $\times$ 10$^{-9}$.
In the chemical reaction network, CS is mainly produced by HCS$^+$, and destructed by H$^+$ and H$_3$$^+$. 
On the other hand, CS$^+$ is mainly produced by reaction of CH with S$^+$, and destructed by electron and H$_2$.
Most calculated abundances of the sulfur-bearing species are larger than their observed values.
The discussion of these discrepancies will be presented later.

\subsubsection{Variation of C/O ratio on the chemistry}
\begin{figure}
\includegraphics{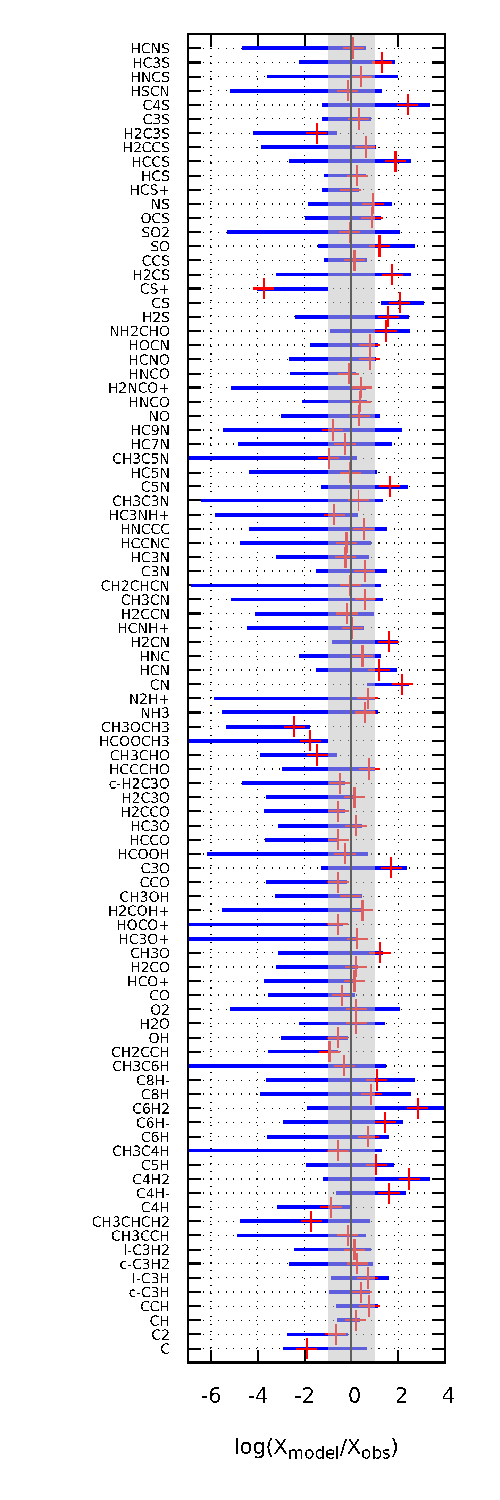}
\caption{Model results for all observed molecules at the TMC-1 CP for the best model 
with an initial carbon abundance of 2.5 $\times$ 10$^{-4}$ and a C/O ratio of 0.5.
The $x$ axis is the logarithmic abundance ratio difference for each observed molecule.
The left endpoint and right endpoint for each molecule is the minimum and maximum value
in the time range of 10$^4$ year to 10$^7$ year respectively. The red cross mark is the time
at the best agreement, i.e.~5.3 $\times$ 10$^5$ year. The vertical grey area between -1 and +1
represent one order of magnitude lower and higher respectively when compared with its observational value.
}
\label{fig.TMC1-CP}
\end{figure}

\begin{figure}
\includegraphics{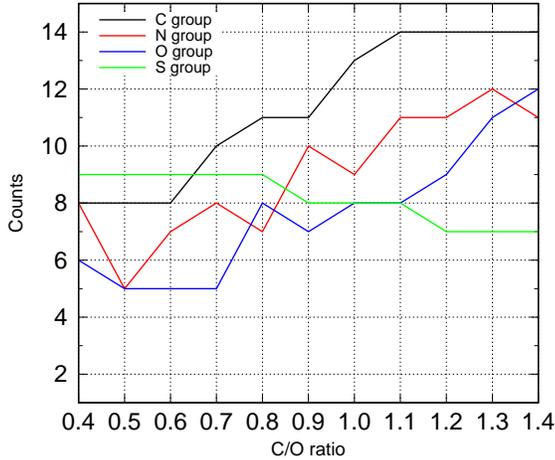}
\caption{The number of disagreement species for the four groups as a function of C/O ratio.
The $y$ axis represents the counted disagreement species when the logarithmic abundance ratio 
differences between models and observations are out of the range from -1 to +1 (see Figure \ref{fig.TMC1-CP}).
The counts are calculated for each model at its own best fit time.
The S group represents all the species that include sulfur element.
The N group represents all nitrogen-containing species with the excluding of S group species.
The O group represents all oxygen-containing species with the excluding of S and N group species.
The C group represents the remaining species, which includes atomic carbon, diatomic carbon and hydrocarbons.
This figure shows the models with an initial carbon abundance of 2.5 $\times$ 10$^{-4}$.
The trends for the four groups are also seen for the low and intermediate initial carbon abundance models.
}
\label{fig.CNOS}
\end{figure}

We noticed that the predicted abundances of S group species are closer to their observed values
when the C/O ratio increases. While the trend for other group species is on the opposite.
The variation of C/O on the chemistry as expressed by the numbers of the disagreement species 
for the four groups can be seen in Figure \ref{fig.CNOS}, which corresponds to the model with an initial carbon
abundance of 2.5 $\times$ 10$^{-4}$.
The trends for the four groups are also presented for the low and intermediate initial carbon abundance models.
The disagreed species are counted when the logarithmic abundance ratio differences between models and observations
are out of the range from -1 to +1 (see Figure \ref{fig.TMC1-CP}).
Note that the counts are calculated for each model at their own best fit time.

\begin{figure}
\includegraphics{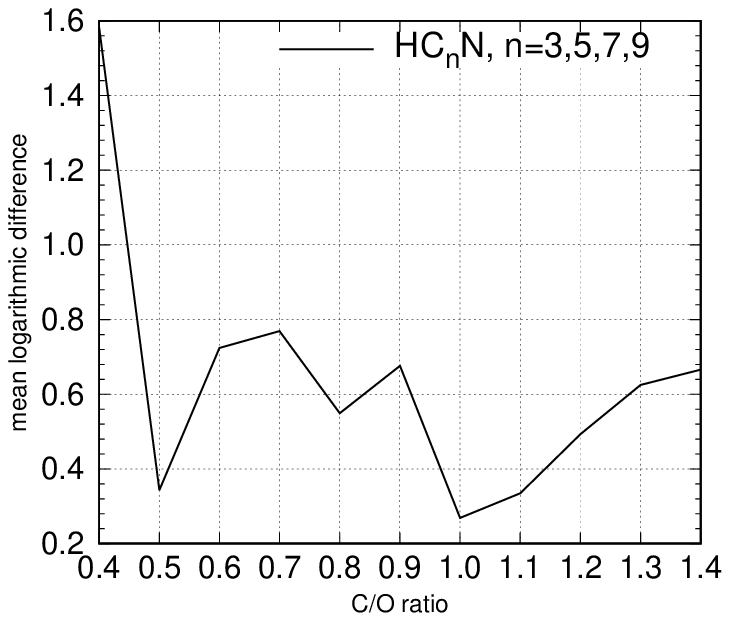}
\caption{The mean logarithmic abundance difference as a function of C/O only for cyanopolyynes 
for the models with an initial carbon abundance of 2.5 $\times$ 10$^{-4}$.
}
\label{fig.HCnN}
\end{figure}

From Figure \ref{fig.CNOS} it can be seen that the C and O group species favor a C/O ratio of 0.5, 
and with the increase of the C/O ratio the disagreement rises highly.
As opposed to the C and O groups, S group species favor a C/O ratio larger than 1.2,
while the overall N group species favor a low C/O ratio similar to the C and O groups.
However, the favored low C/O ratio for the overall N group species 
seems contradiction with the result of \citet{Maffucci2018}, which shows that a C/O ratio of 1.2
can better reproduce cyanopolyynes. We plot the mean logarithmic abundance difference for cyanopolyynes in Figure
\ref{fig.HCnN} using Equation \ref{eq.rms} with $N_{obs}$ only include 
HC$_n$N, where n = 3, 5, 7, 9 at the time determined from the total observed species, 
i.e.~the same evolutionary time used in the Figure \ref{fig.CNOS}.
It can be seen that the minimum value is located at C/O = 1.0, which
is close to 1.2, while our best fit model of C/O = 0.5 is in another minimum location.
Generally, as concluded by \citet{Maffucci2018}, the increase of C/O enhances the production of cyanopolyynes.

Chemically speaking, at the carbon-poor, oxygen-rich condition (C/O $<$ 1), 
most of carbon are locked with oxygen either in the gas phase or on the grain surface 
mainly in the form of CO, CO$_2$, and H$_2$CO.
While at the carbon-rich, oxygen-poor condition (C/O $>$ 1), part of carbon-bearing molecules is hydrogenated, 
thus both in the gas phase and on the grain surface, C$_n$H$_m$ and HC$_n$N are overproduced, 
where n ranges from 3 to 9, and m = 2, 3, 4.
The initial nitrogen is in the neutral atomic form and less abundant than carbon.
It first reacts with hydrocarbons to form nitriles. Most of species in the N group are nitrile-containing
molecules, thus the chemistry of N group species is also affected by the variation of C/O ratio 
and shows similar trend as with the C and O group species.
However, for the subset of HC$_n$N with n = 3, 5, 7, 9 in the N group, 
their abundances could be increased by the overproduced large hydrocarbons when C/O $>$ 1, 
thus show a better fit with the observations than that at the conditions of C/O $<$ 1.
The effect of the variation of C/O on the chemistry of sulfur-containing species in the S group is more complex 
as the elemental compositions of S group species are largely different from each other, 
which indicates that the reactions related to S group species are well connected with other group species.
From Figure \ref{fig.CNOS} it can be seen that 
the difference between the maximum and minimum counts for S group species is only 2, 
which seems not significant when compared with other groups.
There are three species, H$_2$CS, H$_2$S and SO, whose abundances approach to their observed values as C/O increased. 
Instead, the abundance of C$_3$S move away from its observed value. 
For H$_2$CS and H$_2$S, they share similar production pathways when C/O ratio changes. 
However, their destruction involves C$^+$ or C so that they overproduced when C/O $<$ 1.
As for SO, it is produced by the reaction between S and OH, which could be affected by the C/O ratio.
Generally, at C/O $<$ 1, most S group species are slightly overproduced; on the other hand, with the increase of
C/O ratio, the abundances of these species tend to approach their observational values at the best fit time.

\subsection{Chemical variations along the transitional regions}
In the above section, by comparison between the model results and the observations in TMC-1 CP, 
we determined the best initial elemental abundances for carbon and oxygen and the best fit time.
In the following subsections, we show the results of the chemical variations under the predictions 
at the best fit time for the transitional regions where A$_V$ ranges from 1 to 20 mag.

\subsubsection{H, H$_2$ and ionization fraction}
\begin{figure}
\includegraphics[scale=0.8]{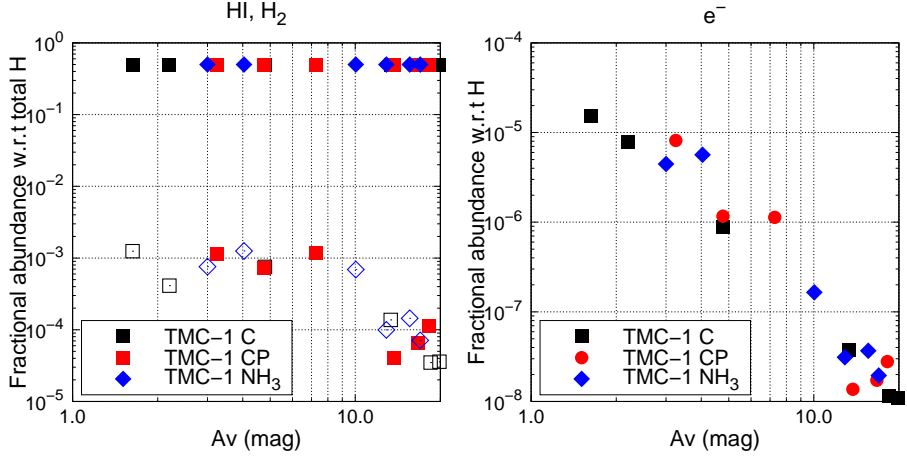}
\caption{The abundances of HI, H$_2$ and e$^-$ as a function of A$_V$.
In the left panel, HI is shown by empty markers, while H$_2$ is shown by filled markers.
Here and after, we use the three emission peaks to represent the three horizontal lines
that across the TMC-1 filament boundary.
}
\label{fig.H}
\end{figure}

As the dominated material, hydrogen is the most abundant element in the ISM both in its atomic form 
and in its molecular form. In the initial condition of the models, we assume all hydrogen exists in its
molecular form. Under this condition, all other species are evolved with time. We determine the best
fit time (i.e~5.3 $\times$ 10$^5$ year) for TMC-1 CP 
by fitting the abundances of all detected species with their predicted values.
At this time, it can be seen from Figure \ref{fig.H} that the corresponding atomic hydrogen abundance at 
A$_V$ $<$ 10 mag is $\sim$10$^{-3}$.
The mean abundance of atomic hydrogen is 10$^{-2.8}$ in TMC and other molecular cloud regions determined by 
HI Narrow Self Absorption (HINSA, \citep{Li2003, Krco2010}).
This consistency between predicted and observed HI abundance suggests that the above assumption is reasonable.
If all hydrogen exists in its atomic form at the initial stage, the conversion timescale for hydrogen to its
molecular form could be longer.

The ionization fraction, or the electron abundance ($x(e^-)$), is a basic physical characteristic of the ISM that 
has a wide influence on the star formation from the driving of the ion-neutral chemistry 
to the dynamics between the gas and magnetic field.
\citet{Goicoechea2009} estimated the gradients of ionization fraction in the Horsehead PDR regions 
from the outer edge to the shielded cores where the value changes from 10$^{-4}$ to 10$^{-9}$ respectively.
Unlike the strong UV field in Horsehead PDR regions (35 in habing unit, \citep{Goicoechea2009}),
the UV field in TMC-1 regions is relatively weak (3.8 in habing uint). There has also
a similar ionization fraction gradient (see the right panel in Figure \ref{fig.H}).
At the edge of TMC-1, the value of $x(e^-)$ is $\sim$10$^{-5}$, 
which is contributed from C, S and cosmic-ray induced ionization of H$_2$.
In the denser part of TMC-1, the ionization fraction is mainly controlled by the cosmic-ray induced ionization
of H$_2$ because UV photons cannot penetrate into the clouds.

\subsubsection{Carbon, Oxygen and their carriers}
\begin{figure}
\includegraphics[scale=0.8]{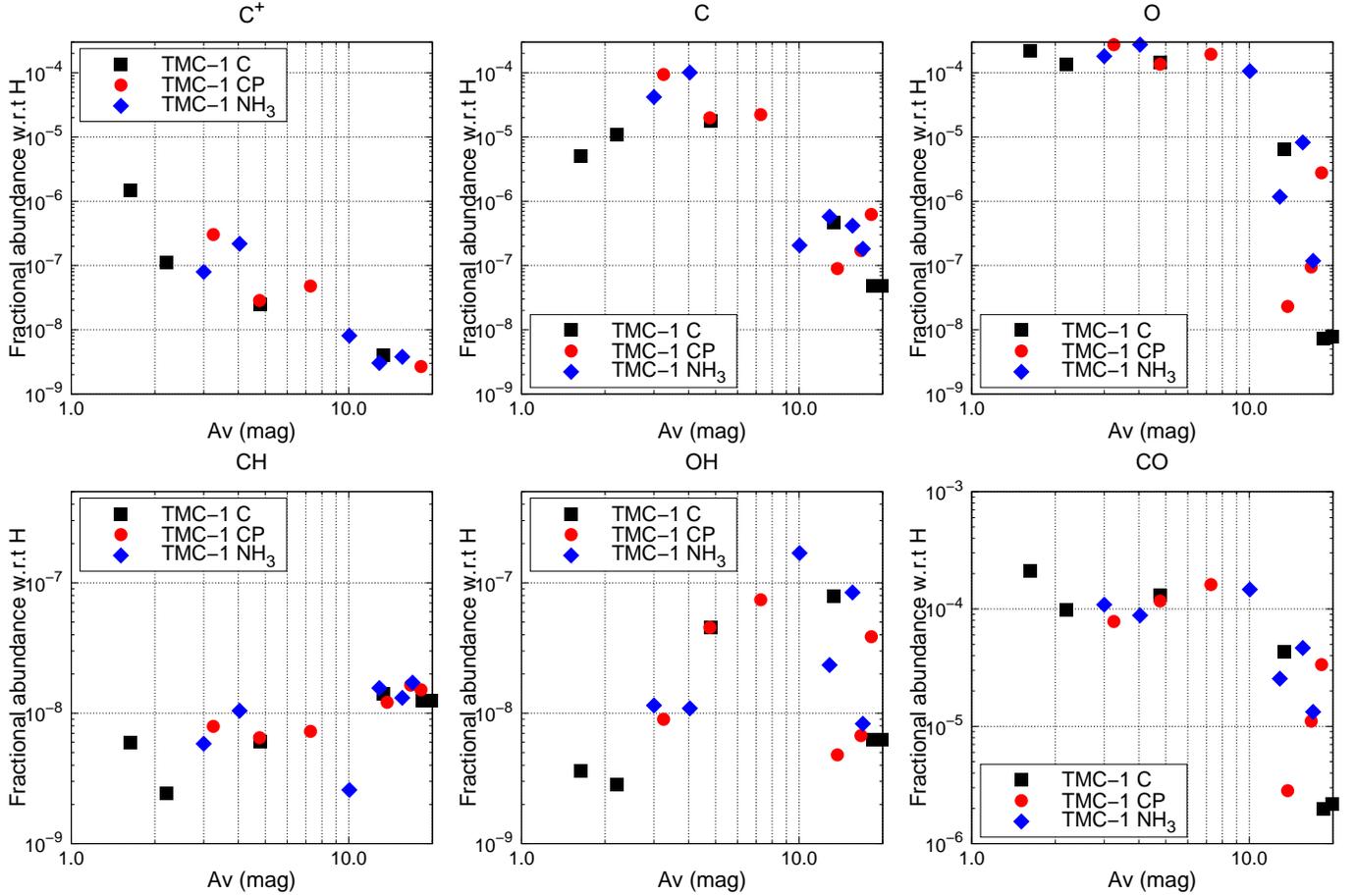}
\caption{The abundances of C$^+$, C and CO as a function of A$_V$.
}
\label{fig.CO}
\end{figure}

Except for the conversion of atomic hydrogen to molecular hydrogen in the transitional regions, 
the species of C$^+$, C and CO are also widely recognized as the evolutionary tracers of
these regions. Figure \ref{fig.CO} shows the calculated abundances of these three species as a function of A$_V$.
It can be seen that the abundance of C$^+$ gradually decreases from the outer low extinction regions
to the inner high extinction regions. 
Meanwhile, C$^+$ are partially converted to C and CO, and the abundance of C peaks around $\sim$4 mag.
The conversion of C$^+$ to CO is efficient because of less strength of UV field in these regions and 
the shielding from H$_2$ and CO itself.
With the continuing increase of A$_V$, eventually most of carbon are locked in the form of CO. 
Meanwhile, at high A$_V$ and density, gas-grain chemistry plays an important role, 
and gas phase CO begins to adsorb onto the grain surface. 
Thus, we see a depletion of CO after A$_V$ $>$ 10 mag. 
The rate of depletion depends on the gas density, with the higher the gas density, the greater the CO depletion rate. 
In the horizontal extensions of TMC-1 CP, it can be seen that at the location of offset 3 from the emission peak,
CO is severely depleted, and this position represents the highest density.
The effect of gas density also has an influence on the other gas phase species as can be seen in the following. 
Molecules that can be easily synthesized in the gas phase adsorb more quickly onto the grains when the gas density is high; 
while for the gas species with low synthesis rate, such as some large molecules, 
their abundances trend to be increased under high density conditions.
In comparison, the densities for the horizontal extensions of TMC-1 C and TMC-1 NH$_3$ are generally
decreased from the inner to the outer region.
It should be noted that our results are presented under the same evolutionary times. 
Some studies have noticed the possible different evolutionary stages 
for the TMC-1 C and TMC-1 CP emission peak \citep{Choi2017, Navarro2021}.
The state of the chemistry could be affected by the different physical conditions and different evolutionary stages of the clouds. 
However, it is difficult to decouple the effects of these two factors on the chemistry 
as the different evolutionary stages may have implied the different physical conditions, 
and it is the focus of our next paper to explore this problem.

The main carriers of oxygen are the atomic oxygen (OI) and CO in the gas phase at A$_V$ $<$ 10 mag.
It can be seen in Figure \ref{fig.CO} that the abundance of atomic oxygen is slightly higher
than that of CO before its depletion.
Recently, \citet{Goldsmith2021} explored the OI emission and absorption in the W3 high-mass star formation regions.
They found that OI is dominated by the foreground absorption of low-excitation atomic oxygen, 
and its abundance could be comparable to, or greater than that of CO.
The highest fractional abundance of OI with respect to H is 6 $\times$ 10$^{-5}$ under 
an initial oxygen abundance of 2.3 $\times$ 10$^{-4}$ in \citet{Goldsmith2021}. 
In our results, the best fit model uses an initial oxygen abundance of 5.0 $\times$ 10$^{-4}$, 
and the abundance of OI could be maintained at a level of $\sim$2.0 $\times$ 10$^{-4}$ when A$_V$ $<$ 10 mag.
Thus, the atomic oxygen, along with C and C$^+$, could be used as the chemical tracers to evaluate
the physical properties of the translucent ISM.

\citet{Xu2016a,Xu2016b} also showed that the primary hydrides CH and OH could be used as the
gas tracers across the TMC boundary.
They derived a fractional abundance (with respect to H$_2$) of 
$\sim$2 $\times$ 10$^{-8}$ and $\sim$2 $\times$ 10$^{-7}$ for CH and OH at A$_V$ = 2 mag, respectively.
Compared with our modeling results, our predictions for the abundances of CH and OH are underproduced
by a factor of 5 and 30, respectively.
\citet{Xu2016a} interpreted the reason for the large discrepancy between the modeled and observational OH abundance
as the shock, which is originated by the compression of the gas in the cloud's edges.
However, the above estimation is under the assumption that all hydrogen is in the form of H$_2$, 
i.e.~the abundance of H$_2$ with respect to the total nuclei is 0.5.
If we assume that a fraction of hydrogen is in the atomic form, 
which is reasonable under the cold neutral medium condition,
then the above discrepancy between our predictions and the observations for the abundances of CH and OH 
could be replaced by a factor of 25 and 5, respectively, when a value of 0.3 
for the abundance of H$_2$ with respect to the total nuclei is used.
It can be seen that the fraction of atomic hydrogen could be important for the primary gas contents
and their chemistry.
Therefore, to better understand the chemistry of CH and OH, direct measurement of primary gas contents,
such as using the HINSA to determine the property of clouds \citep{Zuo2018}, could be helpful.

\subsubsection{S$^+$, S and main sulfur-containing carriers}
\begin{figure}
\includegraphics[scale=0.8]{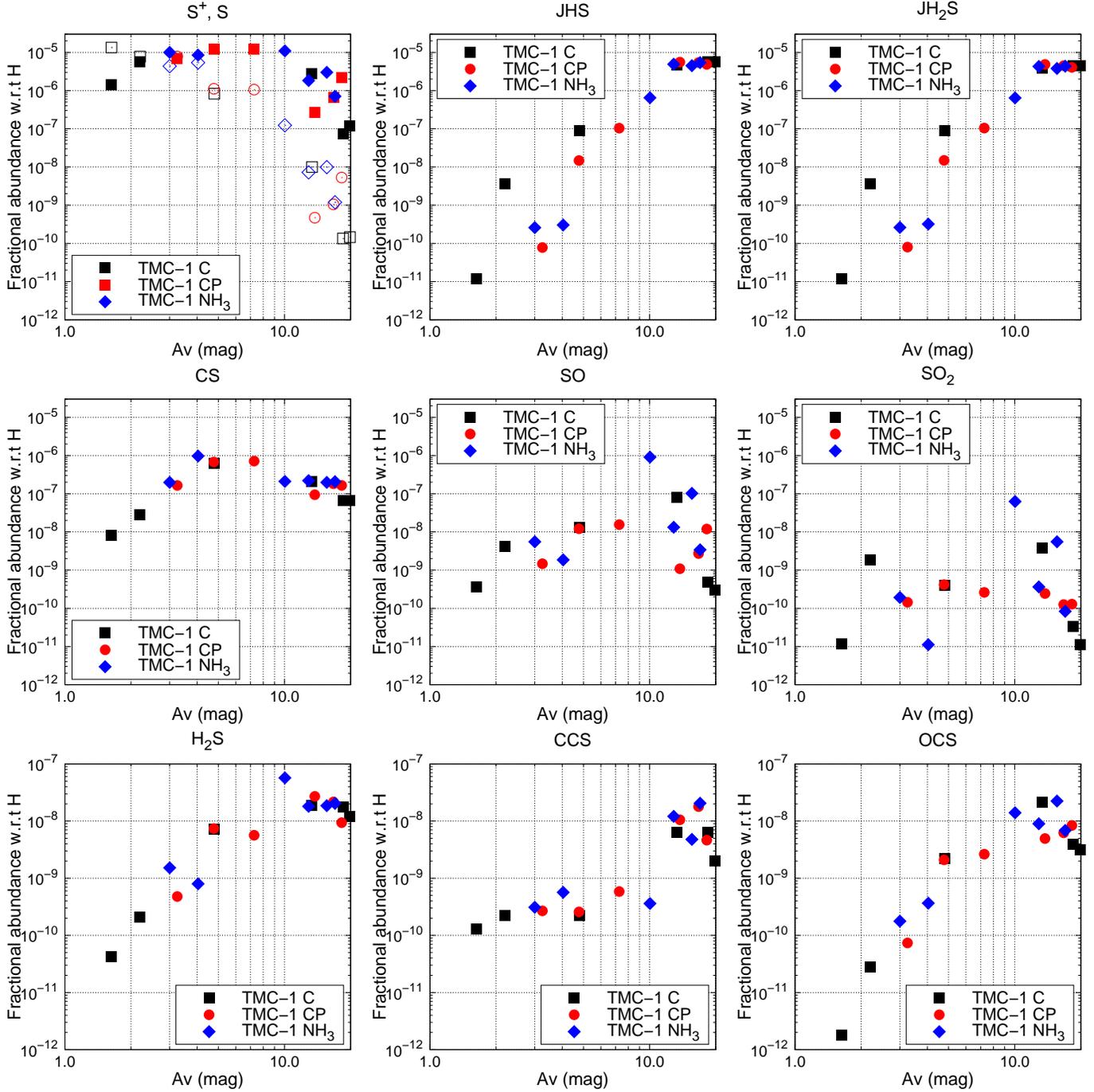}
\caption{The abundances of S$^+$, S and other S-bearing species in the gas surface and on the grain surface. 
In the top left panel, S$^+$ is shown by empty markers, while S is shown by filled markers.
For JHS and JH$_2$S, the prefix ``J'' stands for grain surface species.
}
\label{fig.S}
\end{figure}

The cosmic value of the sulfur elemental abundance is used so that we could study the 
sulfur chemistry across a wide range of A$_V$. 
We demonstrate in Figure \ref{fig.S} the calculated most abundant gas phase and grain surface S-bearing molecules. 
Similar to the trends of C$^+$ and C as a function of A$_V$, S$^+$ (shown in empty markers) remain its
ionic atomic form when A$_V$ $<$ 3 mag, and are converted to neutral atomic form (shown in filled markers) 
at higher A$_V$. When A$_V$ $>$ 10 mag, sulfur is mostly adsorbed onto the grains and 
hydrogenated to JHS and JH$_2$S subsequently (hereafter, ``J'' stands for grain surface species) 
as shown in the upper panels of Figure \ref{fig.S}. For example, at the TMC-1 CP, 
which corresponding to an A$_V$ of 18.2 mag, 59 percent of the total sulfur abundance
is in the form of JHS and JH$_2$S, and 20 percent is occupied by other grain surface species such as
JHSCN, JHSO, JOCS and JNS, whereas 15 percent by the gas phase neutral atomic sulfur, 
and the remaining small amount is occupied by other S-bearing species. 
For the gas phase S-bearing species, they are mainly produced via the gas phase reactions with the exception
of H$_2$S, which could be produced by the non-thermal desorption of grain surface JH$_2$S.
Among them, CS, SO and SO$_2$ reach their peak abundances when A$_V$ are around or less than 10 mag, 
while H$_2$S, CCS and OCS reach their peak abundances in the inner part of the cloud, which 
has large A$_V$ and density.
Moreover, it can also be seen in Figure \ref{fig.S} that the abundances of SO and SO$_2$ are more sensitive 
to the physical conditions at certain A$_V$, while other species are much less affected.

Compared with other abundant gas phase tracers, such as CO, there is a lack of 
abundant S-bearing molecular tracers in the gas phase for the sulfur chemistry.
The abundance of CS reaches 10$^{-6}$ in the hot corino condition \citep{Codella2021},
and depletes to 10$^{-8}$ in the hot core and dark cloud \citep{Luo2019, Bulut2021}.
In Figure \ref{fig.S}, it can be seen that the CS abundance could reach the highest value of 10$^{-6}$ at A$_V$ = 4 mag.
However, the dominant sulfur-carriers above this level are atomic sulfur in the gas phase and JHS and JH$_2$S on the grain surface.
The reason is twofold: the initial elemental S abundance is less than that for C, O and N, 
and S is heavier than C, O and N, making it easier to adsorb on the grains.

\subsubsection{Nitrogen-bearing molecules}
\begin{figure}
\includegraphics[scale=0.8]{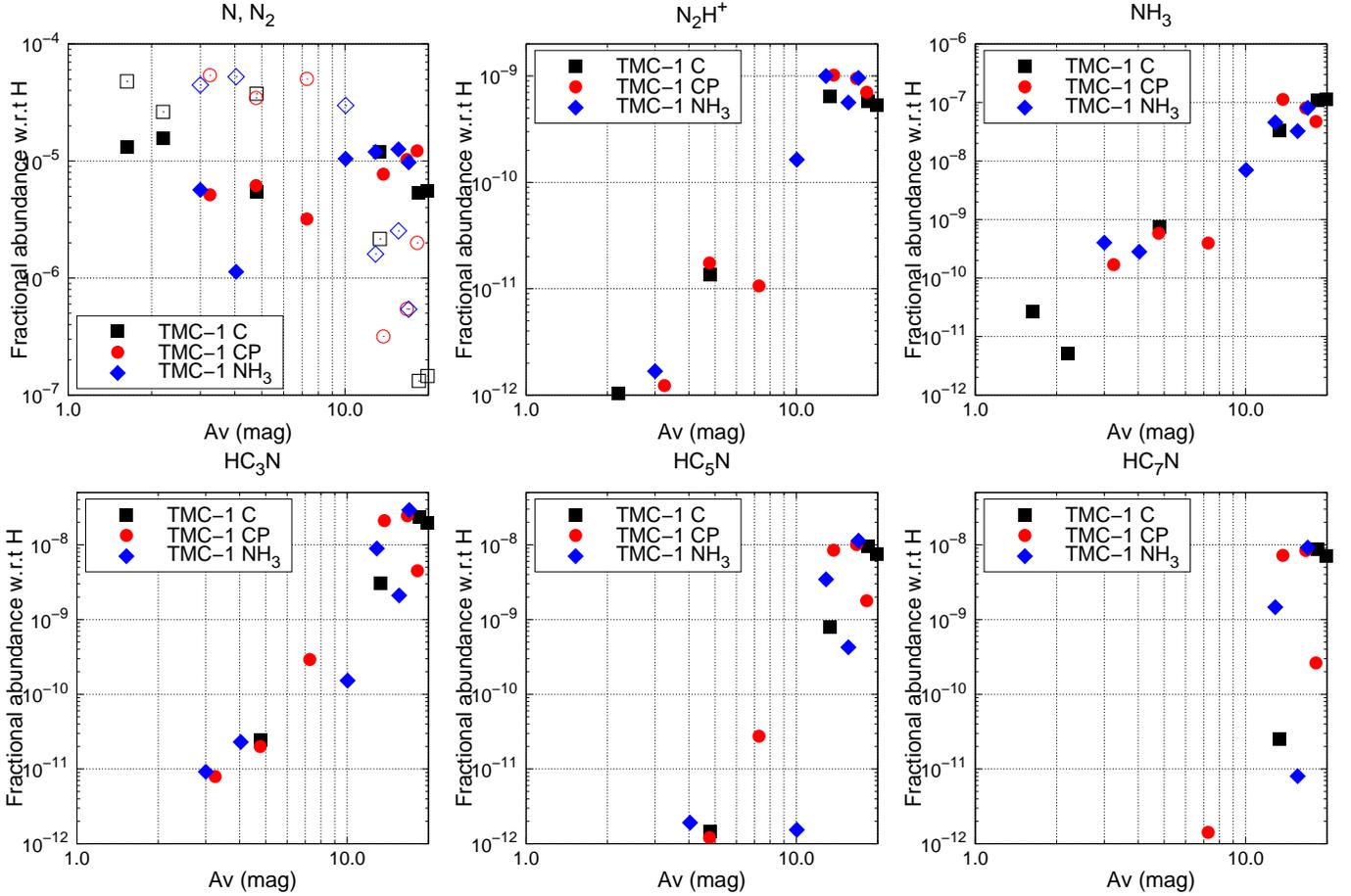}
\caption{The gas phase abundances of N, N$_2$ and other nitrogen-bearing species as a function of A$_V$. 
In the top left panel, N is shown by empty markers, while N$_2$ is shown by filled markers.
}
\label{fig.N}
\end{figure}

Figure \ref{fig.N} shows the gas phase N (in empty markers), N$_2$ (in filled markers) 
and other nitrogen-bearing species as a function of A$_V$.
The initial nitrogen is in the neutral atomic form due to its slightly higher ionization potential (14.5 eV)
than that of H atom. It can be seen in Figure \ref{fig.N} that the majority of nitrogen remains in its neutral atomic
form when A$_V$ is less than 10 mag, and mostly convert to N$_2$ at higher A$_V$.
N$_2$ is not observable at infrared or millimeter-wavelength transitions, but can be estimated by the 
main production channel of N$_2$H$^+$: N$_2$ + H$_{3}^{+}$ $\rightarrow$ H$_2$ + N$_2$H$^+$.
\citet{Womack1992} estimated a gas phase N$_2$ abundance of 4 $\times$ 10$^{-6}$ through N$_2$H$^+$ 
under the dark cloud conditions, and suggested that gas phase N$_2$ as the dominant nitrogen-bearing species. 
In our models, for A$_V$ $>$ 10 mag, the N$_2$ abundances varies between 5 $\times$ 10$^{-6}$ and 
1 $\times$ 10$^{-5}$, which is less than 3 times of the observational value.
N$_2$ is mainly produced by the reaction of N + CN $\rightarrow$ C + N$_2$ at earlier time 
($\sim$5 $\times$ 10$^4$ year), and by the cosmic-ray desorption of grain surface JN$_2$ at later time.
Two of the destruction pathways of N$_2$ are the reaction of N$_2$ + H$_{3}^{+}$ $\rightarrow$ H$_2$ + N$_2$H$^+$
and the adsorption onto the grain surface.
The production of NH$_3$ is also starting from N$_2$. 
Firstly, N$^+$ is produced by the reaction of N$_2$ + He$^+$ $\rightarrow$ N$^+$ + N + He. 
Then, following the sequence of N$^+$, NH$^+$, NH$_{2}^{+}$ and NH$_{3}^{+}$ with the 
reaction of H$_2$, NH$_{4}^{+}$ is produced. 
Finally, NH$_3$ is produced by the reaction of NH$_{4}^{+}$ + e$^-$ $\rightarrow$ H + NH$_3$.
The family of cyanopolyynes, HC$_n$N (n = 3, 5, 7, 9) are produced by the reaction of CN with hydrocarbons 
of C$_n$H$_2$ (n = 2, 4, 6, 8). 
Overall, except for N$_2$, it can be seen from Figure \ref{fig.N} that the abundances of most nitrogen-bearing 
species gradually increase with larger A$_V$ and peak at the highest density.
Generally, the nitrogen-bearing species trace the regions with higher A$_V$ and larger n$_H$.

By comparing the gas phase N$_2$ in Figure \ref{fig.N} with the grain surface JN$_2$ in the following Figure \ref{fig.JH2O}, 
it can be seen that the trends for N$_2$ and JN$_2$ are very similar, both having a turning point
around A$_V$ of 4--5 mag.
The grain surface JN$_2$ is mainly produced by the grain surface reaction of JN + JNO $\rightarrow$ 
JO + JN$_2$, and also by the gas phase N$_2$ adsorption. The major consuming channel for JN$_2$ is its cosmic-ray 
desorption to the gas phase, which is independent of the A$_V$. Thus, the gas phase N$_2$ and grain surface JN$_2$ 
are well coupled through the adsorption and desorption mechanism.
By comparing the estimated N$_2$ abundances in the dark clouds and in the warm clouds, which shows no difference, 
\citet{Womack1992} suggested that dust grains may not play an important role in the N$_2$ chemistry.
However, according to our above analysis, 
we suggest that dust grains indeed could play an important role in the N$_2$ chemistry 
by keeping N$_2$ on the grain surface and releasing back to the gas phase at a later time.

\subsubsection{Complex organic molecules}
Complex organic molecules (COMs), which are defined as carbon-containing molecules with at least six atoms \citep{Herbst2009},
represent the advanced complexity during the development of the chemistry. 
Some COMs, such as CH$_3$OH and CH$_3$CHO, have been detected 
in the diffuse clouds toward Galactic center \citep{Thiel2017} and in the translucent clouds \citep{Turner1998, Turner1999}.
\citet{Thiel2017} estimated the abundances with respect to H$_2$ for CH$_3$OH, CH$_3$CHO and NH$_2$CHO 
in the diffuse clouds toward the line-of-sight of Sagittarius B2 of 8.1 $\times$ 10$^{-8}$, 
$<$3.2 $\times$ 10$^{-9}$ and $<$8.5 $\times$ 10$^{-10}$, respectively at a $\upsilon_{LSR}$ of 9.4 km/s.
While the fractional abundances with respect to H$_2$ for CH$_3$OH and CH$_3$CHO 
in translucent clouds are in the range of 10$^{-9}$ to 10$^{-8}$ \citep{Turner1998, Turner1999}.
However, the origin of these COMs in the diffuse and translucent clouds is still in debate.
\citet{Liszt2018} argued that the nature of the host gas environment could complicate this situation.
Indeed, as suggested in \citet{Thiel2017}, the diffuse clouds chemistry could be enriched
by the mixing between the dense part and diffuse part of the gas.

\begin{figure}
\includegraphics[scale=0.8]{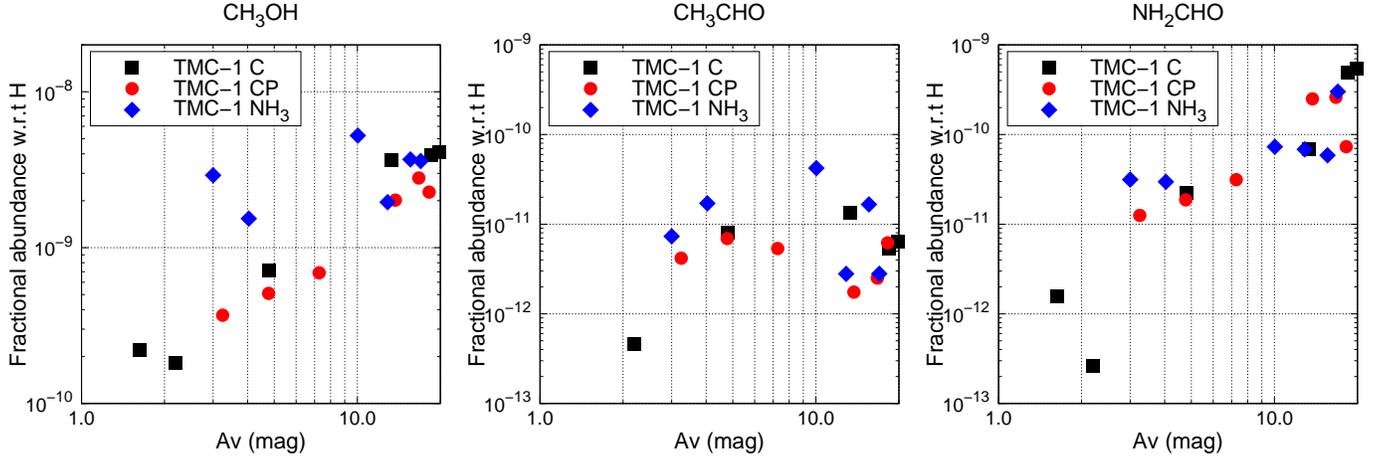}
\caption{The gas phase abundances of three complex organic molecules as a function of A$_V$. 
}
\label{fig.COMs}
\end{figure}

Figure \ref{fig.COMs} shows the modelled abundances of the three COMs as a function of A$_V$.
At intermediate A$_V$ (around 5 mag), the modeling CH$_3$OH abundance is consistent with the observations, 
while CH$_3$CHO and NH$_2$CHO are underproduced.
CH$_3$OH is first synthesized on the grain surface and then desorbed into the gas phase by the chemical process. 
The main chemical process for the production of gas phase CH$_3$OH is the chemical desorption of the 
exothermic surface reactions. Under a grain temperature of $\sim$10 K, the thermal desorption is not efficient, 
and other desorption mechanisms, such as the cosmic-ray desorption, also contribute less 
due to the high desorption energy of grain surface methanol.
The origin of gas phase NH$_2$CHO is also in a similar way. 
However, CH$_3$CHO is mainly produced by the gas phase reactions of O + C$_3$H$_7$ $\rightarrow$ CH$_3$ + CH$_3$CHO 
and O + C$_2$H$_5$ $\rightarrow$ H + CH$_3$CHO. 
The destruction pathways for these COMs are mostly by the abundant atomic carbon and by the UV photodissociation.
The high radiation field in the diffuse and translucent environment is one of the reasons 
for the low abundances of the COMs. On the one hand, COMs could be dissociated by the high energy photons.
On the other hand, the grain surface chemistry, which is one of the channels that produce the formation of COMs,
could also be suppressed by the high radiation.
Nevertheless, our modeling results show that COMs indeed could be produced 
at an abundance of higher than 10$^{-11}$ in the translucent clouds with A$_V$ down to 3 mag.

\subsubsection{Variation of grain ice composition}
\begin{figure}
\includegraphics[scale=0.8]{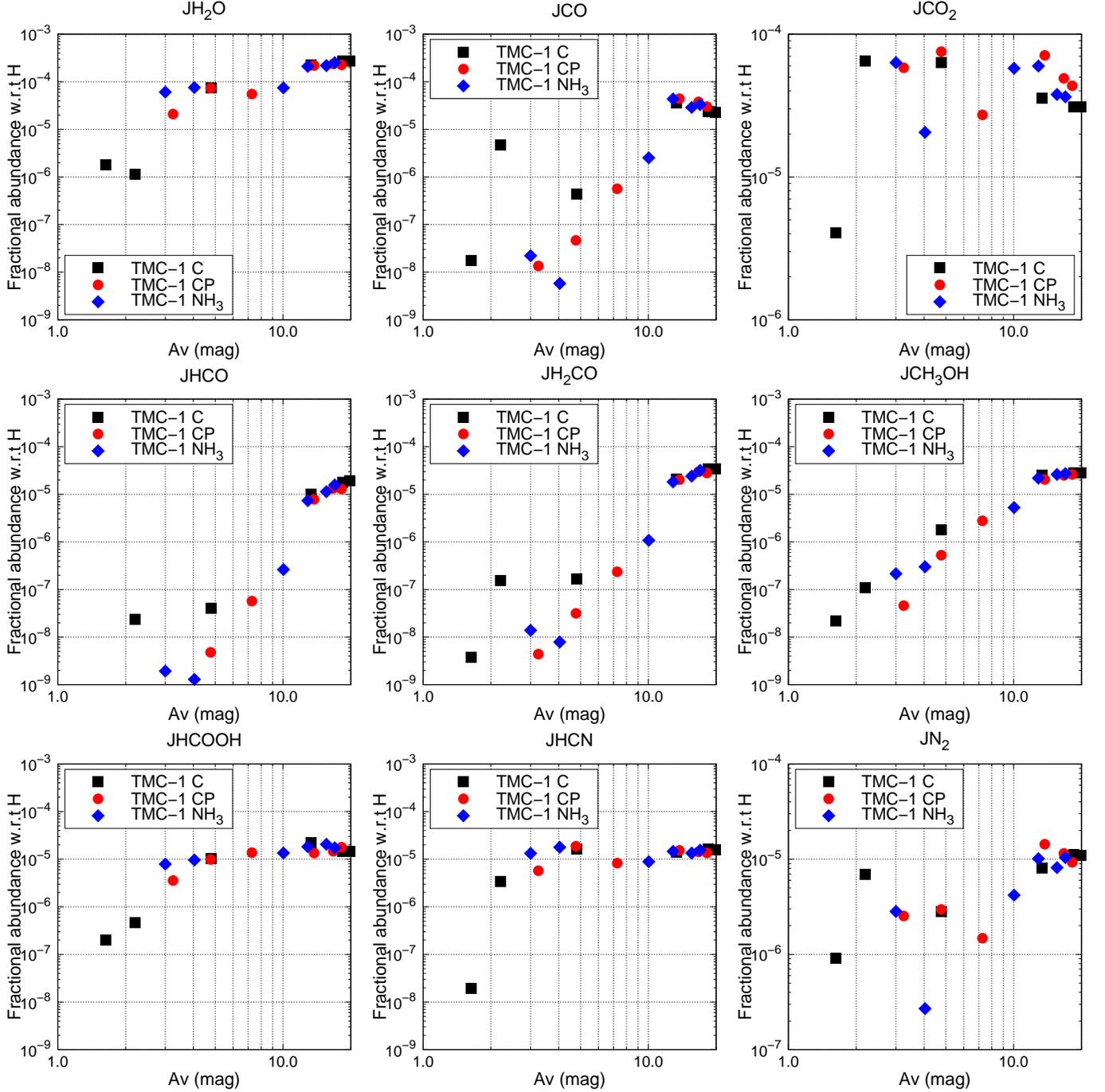}
\caption{The abundances of grain surface species as a function of A$_V$. 
The prefix ``J'' stands for grain surface species.
}
\label{fig.JH2O}
\end{figure}

Water ice is widely present in the dense clouds based on the observations toward the 
line-of-sight dense part of clouds toward the background stars \citep{Gibb2000,Gibb2004}.
However, there was no strong evidence for the existence of solid-state water in the 
diffuse-dense transition region \citep{Poteet2015}. Recently, \citet{Potapov2020} shows that 
water ice could be expected in the diffuse ISM by the laboratory experiments.
Although the mixture of grain ice mantle compositions is hard to determine,
it is interesting to investigate the variation of grain ice compositions under different conditions 
for the exploration of grain surface chemistry and the role of gas-grain interactions.
In Figure \ref{fig.JH2O} we show the predicted abundances of grain surface species as a function of A$_V$.
Water ice (JH$_2$O) is the most abundant species when A$_V$ is higher than 4 mag. However, most JH$_2$O are dissociated
by UV photons when A$_V$ is less than 4 mag, thereby making its decrease in the grain ice mantles.
Under this condition, the three most abundant ice compositions are JCO$_2$, JCO and JN$_2$.
Among them, JCO$_2$ are the dominant ice composition at low A$_V$ because of the reaction between JCO and JOH
and between JO and JHOCO, where JOH is provided by the photodissociation of JH$_2$O.
The abundance of JCO has a turning point around 4--5 mag resulting from the competition 
between its destruction and formation pathways. 
The destruction pathways of JCO are the reactions with JH and JOH, 
while the dominant formation pathways of JCO are the UV photon dissociation and cosmic-ray induced UV photon 
dissociation of grain surface JCO$_2$. Note that when A$_V$ is less than 5 mag, the contribution of the gas phase 
CO adsorption is unimportant. When A$_V$ is smaller than 5 mag, the abundance of JCO 
gradually increased because of the increased strength of UV photon and cosmic-ray induced UV photon dissociation of JCO$_2$.
On the contrary, starting from the turning point, with the increase of A$_V$ the effect of photodissociation of JCO$_2$ 
becomes weaker and gas phase CO adsorption becomes more important, thus leading to the increase of JCO abundance.
After A$_V$ $>$ 10 mag, part of JCO is hydrogenated to produce JHCO, JH$_2$CO and JCH$_3$OH gradually.
It can be seen in Figure \ref{fig.JH2O} that the abundances of JHCO, JH$_2$CO and JCH$_3$OH are more dependent on A$_V$ 
and reach their highest values at the largest A$_V$ considered in the models.
On the other hand, JH$_2$O, JCO$_2$, JHCOOH, JHCN and JN$_2$ are generally kept constant when A$_V$ is higher than 4 mag.
JHCN and JN$_2$ are produced by the grain surface reaction of JH$_2$ + JCN $\rightarrow$ JH + JHCN and 
JN + JNO $\rightarrow$ JO + JN$_2$, respectively. They are both the dominant nitrogen-bearing molecules 
on the grain surface.

\begin{figure}
\includegraphics[scale=0.8]{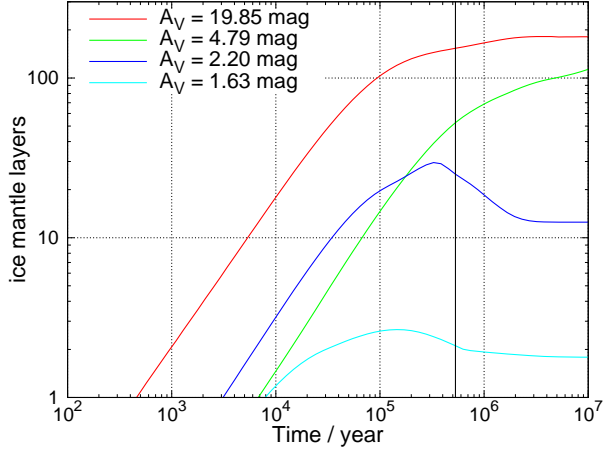}
\caption{The development of the grain ice mantle layers as a function of time for different A$_V$.
The vertical black line represent the best fit time of 5.3 $\times$ 10$^5$ year.
}
\label{fig.iceLayer}
\end{figure}

The grain ice mantle composition depends on many factors, such as the radiation field and n$_H$ density.
We show in Figure \ref{fig.iceLayer} the development of the grain ice mantle layers as a function of time
in four positions at the horizontal line of TMC-1 C and its offset 3, 4 and 5 (see Table \ref{tab.para}).
It can be seen that the ice layers gradually decrease with the decrease of A$_V$.
The ice layers for A$_V$ of 2.2 mag are thicker than that for A$_V$ of 4.8 mag at the time before 10$^5$ year.
This is due to two times higher of n$_H$ density toward the position at A$_V$ of 2.2 mag than 
that toward the position at A$_V$ of 4.8 mag.
We list in Table \ref{tab.JH2O} the most abundant fractional ice composition at these positions.
In comparison, we also list the observational values
of the fractional ice comparison toward Elias 16 where the location is in TMC and 
the dust-embedded young stellar object W33A \citep{Gibb2000}.
Water ice is the dominated ice composition at the dense part of the clouds.
CO and CO$_2$ ice show comparable fractions in the ice mantles.
Meanwhile, the hydrogenated H$_2$CO, CH$_3$OH, NH$_3$ and CH$_4$ ingredients increased the ice composition complex.
However, in the diffuse-translucent part of the clouds, with the effect of the changes of physical conditions,
such as radiation field and n$_H$ density on the grain ice mantles, the grain ice compositions are greatly altered. 
Under these conditions, the CO$_2$ ice becomes the dominant ice composition, 
with H$_2$O, N$_2$ and CO ice as the main ingredients.

\begin{table}
\caption{Fractional ice composition for different A$_V$ and different sources.}
\label{tab.JH2O}
\begin{tabular}{l|cccc|cc}
\hline \hline
            &              & TMC-1 C       &         &           &               &       \\
Species     & 1.63 mag$^a$ & 2.2 mag$^a$   & 4.8 mag & 19.9 mag  & Elias 16$^b$  & W33A$^c$  \\
\hline  
JH$_2$O     & 45           & 2             & 100     & 100       & 100           & 100   \\
JCO         & 0 $^d$       & 7             & 1       & 8         & 25            & 8     \\
JCO$_2$     & 100          & 100           & 85      & 11        & 24            & 13    \\
JHCO        & 0            & 0             & 0       & 7         & -$^e$         & -     \\
JH$_2$CO    & 0            & 0             & 0       & 13        & -             & 6     \\
JCH$_3$OH   & 1            & 0             & 2       & 10        & $<$ 3         & 18    \\
JHCOOH      & 5            & 1             & 14      & 5         & -             & 7     \\
JXCN$^f$    & 1            & 2             & 2       & 3         & $<$ 1.5       & 3.5   \\
JNH$_3$     & 0            & 0             & 4       & 4         & $<$ 10        & 15    \\
JCH$_4$     & 0            & 0             & 0       & 1         & -             & 1.5   \\
JHCN        & 0            & 5             & 22      & 6         & -             & -     \\
JN$_2$      & 22           & 11            & 4       & 4         & -             & -     \\
\hline
\end{tabular}
\\

Notes.\\
The prefix ``J'' of a molecule stands for grain surface species. All values are interpreted as percentages.\\
$^a$ Percentages with respect to JCO$_2$ for this column, 
while percentages with respect to JH$_2$O for all other columns.\\
$^b$ Cited from \citet{Gibb2004}.\\
$^c$ Cited from \citet{Gibb2000}.\\
$^d$ ``0'' means less than 1 percent.\\
$^e$ ``-'' means undetected.\\
$^f$ The main carrier for JXCN is most probably a cyanate or the OCN$^-$ ion \citep{Gibb2004}. Here we assume
JHOCN as the main carrier for JXCN.
\end{table}

\section{Discussion}\label{sec.discuss}
In this section, we discuss the discrepancy 
between the modeling results of large hydrocarbons and sulfur-bearing molecules from their observations.

It can be seen in Figure \ref{fig.TMC1-CP} that the hydrocarbons with the number of carbon fewer than 3 are well
produced, while the large hydrocarbons for the family of C$_n$H$^-$ and C$_n$H$_2$ with n = 4, 6, 8, are overproduced.
The anions are mainly produced by the electron attachment with a neutral species, such as C$_n$H, 
and are destructed by hydrogen atom addition. 
The two reaction pathways finally lead to the formation of C$_n$H$_2$. 
and their rates are rapid under laboratory measurements \citep{Agundez2013}.
It is notable that the abundance of atomic carbon is underproduced by almost two orders of magnitude, 
which means that too much carbon is converted to other carbon-bearing species.
The initial donor of carbon is C$^+$, and hydrocarbons grow by the reactions of C$^+$ or C with 
C$_{n-1}$H$_2$ or C$_{n-1}$H$_2^{+}$ respectively.
As the result, the overproduced large hydrocarbons may indicate the too efficient conversion processes,
although we cannot rule out the possibility of the inefficient destruction pathways of large hydrocarbons.
In the reaction network, hydrocarbons are mainly dissociated by O or H$_{3}^{+}$, or react with C to form larger hydrocarbons. 
There are also charge exchange reactions in the gas phase, but their effect on the decrease of hydrocarbons is minor. 
To the best of our knowledge, 
the charge exchange reactions of hydrocarbons with grains are not well considered in the network, 
thus their effect on the destruction of hydrocarbons is hard to determine.

The sulfur chemistry is another unsolved problem both in terms of observation and astrochemical modeling.
The cosmic value of the sulfur elemental abundance is about 1.5 $\times$ 10$^{-5}$. 
While for the commonly detected sulfur-bearing molecules (such as CS, CCS, SO and H$_2$S, etc.) 
in dense clouds and other environments, such as diffuse clouds and infrared dark clouds, 
their abundances are generally on the level of 10$^{-9}$ or below 
\citep{Neufeld2015, Fuente2019, Navarro2020, Xie2021}.
It is also considered that most sulfur could deplete onto the dust grains.
Observationally, the abundances of solid H$_2$S, SO$_2$ and OCS ice with respect to H$_2$O ice 
could reach 0.7, 0.8 and 0.04 percent, respectively \citep{Jimenez2011, Boogert1997, Palumbo1995}.
In the dense part of the clouds, our modeling results show that 
the dominated carrier for sulfur is atomic sulfur in the gas phase (10$^{-6}$--10$^{-7}$ with respect to hydrogen nuclei)
and solid HS and H$_2$S on the ice mantles (2 percent with respect to water ice).

As presented in Section \ref{sec.fitmodel}, our model could reproduce 71 percent of 93 observed species in TMC-1 CP 
where 9 out of 20 sulfur-bearing species are out of one-order-of-magnitude criterion.
\citet{Vidal2017} reproduced a similar percentage with a total observed species of 61 
in which 11 sulfur-bearing molecules are included. 
To solve the overproduced CS problem, \citet{Bulut2021} considered the reaction 
CS + O $\rightarrow$ CO + S with newly theoretical calculated reaction rate. However, their modeling
result for the abundance of CS is still overproduced.
As can be seen in Figure \ref{fig.TMC1-CP}, most overproduced species are carbon-sulfur chains.
In the chemical reaction network, sulfur chemistry is well connected with the hydrocarbons, 
such as the reactions with C$_n$H and C$_n$H$_2$ family \citep{Agundez2013}.
This indicates that the overproduced sulfur-bearing species and hydrocarbons may connect with each other.
However, the existence of different types of interstellar dust grains could complicate the situation. 
Unlike the silicate dust grain used in our model, 
there are also carbonaceous grains and polycyclic aromatic hydrocarbons (PAHs) \citep{Draine2003}. 
The destruction of carbonaceous grains could provide an additional source of carbon elements 
that may eventually influence the chemistry. 
And the inclusion of PAHs in the chemistry has been shown 
to improve the agreement with observations for some species \citep{Wakelam2008,Ge2020}.
Further explorations on the puzzling sulfur chemistry and the link between sulfur and hydrocarbons may need
to better understand their formation and destruction pathways in the dark clouds.

Last but not least, as summarized in the Introduction section, the large number of emerged molecules 
detected in TMC-1 recently challenged the modeling of astrochemistry. 
Due to the lack of most of these new species in the reaction network, 
a comprehensive understanding of their evolution is less constrained, 
and indeed current simple modeling of their formation tends to underproduce their abundances.
Thus, a systemic census on the formation and destruction pathways for these newly emerged species
will be carried out in the further studies.

\section{Summary}\label{sec.conc}
In this paper, we investigated the chemical variations across the TMC-1 filament from the outer 
regions to the innermost regions, which represent the transitions from the translucent phase 
to the dense phase as indicated by the extinction. Three horizontal lines were chosen starting 
from three emission peaks, which are the cyanopolyynes emission peak (TMC-1 CP), the ammonia 
emission peak (TMC-1 NH$_3$) and the dust continuum peak (TMC-1 C). Their physical parameters were 
extracted from \citet{Fuente2019}, which were used to constrain the physical conditions of our 
studied regions. Among them, TMC-1 CP is the frequently observational target in the literature 
where more than 90 molecules have been identified. 
We divided the identified molecules into four groups by the representative species, which are C group 
(hydrocarbons), S group (all the sulfur-containing molecules), N group (nitrile-containing molecules) 
and O group (the remaining oxygen-containing molecules).
Under the above conditions, we first constrained our key chemical parameter, 
(i.e.~the initial elemental abundances of carbon and oxygen) by comparison between the observational 
molecular abundances and the predicted modeling abundances, and determined the evolutionary chemical 
timescale of TMC-1 CP. 
We then studied the chemical variations of different groups of molecules as a function of extinction, 
and investigated the most common and abundant grain ice mantle compositions. 
These results could serve as the template for the study of the transitional regions from 
the translucent phase to the dense phase of the molecular clouds. 
Finally, we presented and analyzed the chemical evolutions of the tracers in these regions, 
especially for the grain surface species that show a distinct difference between the translucent phase and the dense phase. 

Our model results can be compared with the observational results from \citet{Fuente2019}. 
The trends of the chemical abundances as a function of A$_V$ for species CO, CS and SO are successfully reproduced. 
While for N$_2$H$^+$, the comparison reaches agreement when A$_V$ $>$ 10 mag. 
\citet{Xu2016b} studied the CH across the edge of the TMC where A$_V$ $<$ 2 mag, 
and showed that its abundance keeps nearly constant across that region. 
Our model shows that the abundance of CH also keeps nearly constant when A$_V$ $>$ 2 mag. 
We conclude that the chemical model is a suitable tool to explore the physicochemical conditions of the clouds. 
We have presented several other species for the four groups, 
which could be useful for future studies of the transitional regions, 
such as the edge of a molecular cloud or the region from the envelope to the inner part of a dense core.
For the ice mantle species, the future observations of James Webb Space Telescope (JWST) 
could be essential for the identifications of these grain ice mantle compositions.

We summarize the main conclusions as follows:
\begin{enumerate}[1.]
\item Our best fit model for TMC-1 CP could fit 71 percent observed molecules within one order of magnitude, 
and shows that the evolutionary time of TMC-1 CP is $\sim$5.3 $\times$ 10$^5$ years, 
which is consistent with the CO depletion timescale found in the molecular cores located 
in the TMC-1 (Figure \ref{fig.rms}, Figure \ref{fig.TMC1-CP}).

\item The variation of initial carbon to oxygen elemental abundance ratio (C/O) has different impact 
on different group of molecules. Specifically, C and O group molecules favor a low C/O condition (C/O = 0.5), 
while S group molecules seem to favor a high C/O condition.
N group molecules are also affected by the variation of C/O. 
Although cyanopolyynes (HC$_n$N, where n = 3, 5, 7, 9) favor the condition at C/O = 1.0,
the overall N group molecules could best fit by a low C/O of 0.5 (Figure \ref{fig.CNOS}, Figure \ref{fig.HCnN}).

\item As the molecular cloud evolves from the translucent phase to dense phase, there is a conversion process for 
carbon-, oxygen-, nitrogen- and sulfur-bearing species, i.e.~from the atomic form to their main reservoir form.
However, the dominated converted form depends on the elements. 
Atomic carbon and oxygen first convert to CO when A$_V$ $\approx$ 2 mag (Figure \ref{fig.CO}). 
When A$_V$ $\approx$ 3--4 mag, ionic atomic sulfur converts to neutral atomic sulfur, which is the main gas phase reservoir. 
Sulfur could also be depleted onto grain surface to form JHS and JH$_2$S (``J'' stands for grain surface species) 
when A$_V$ is larger than 10 mag (Figure \ref{fig.S}). 
The conversion of atomic nitrogen to N$_2$ begins when A$_V$ is larger than 10 mag (Figure \ref{fig.N}).

\item COMs, such as CH$_3$OH, CH$_3$CHO and NH$_2$CHO, 
could also exist in the translucent gas for A$_V$ down to 3 mag (Figure \ref{fig.COMs}).
The main chemical mechanism for the production of CH$_3$OH and NH$_2$CHO is the chemical desorption of 
the grain surface reactions. While for CH$_3$CHO, it is directly produced in the gas phase.

\item Grain ice mantle composition has a dependence on A$_V$. 
H$_2$O ice is the dominant ice composition at A$_V$ $>$ 4 mag. 
CO ice is severely decreased around A$_V$ of 4--5 mag. 
The composition of CO$_2$ ice keeps nearly constant at A$_V$ $>$ 4 mag 
and is the dominant component when A$_V$ $<$ 4 mag. 
N$_2$ ice tends to enrichment at low extinction condition 
(Figure \ref{fig.JH2O}, Table \ref{tab.JH2O}).

\end{enumerate}

\acknowledgments
We thank the reviewer's helpful comments on improving the manuscript.
This work is supported by the National Natural Science Foundation of China (NSFC) grant No.~11988101 and No.~11725313.
D.Q.~acknowledges supports from the NSFC grant No.~11973075.
X.H.Li acknowledges the Xinjiang Tianchi project (2019).
L.X.~acknowledges supports from the science and technology projects in Jilin Province Department of Education grant JJKH20200607KJ,
and from the NSFC grant No.~11703075, CAS “Light of West China” Program 2017-XBQNXZ-B-020 
and the Heaven Lake Hundred-Talent Program of Xinjiang Uygur Autonomous Region of China.

\bibliographystyle{aasjournal}

\begin{thebibliography}{99}

\bibitem[\protect\citeauthoryear{Navarro-Almaida et al.}{2020}]{Navarro2020} 
Navarro-Almaida D., Le Gal R., Fuente A., Rivi{\`e}re-Marichalar P., Wakelam V., Cazaux S., Caselli P., et al., 2020, A\&A, 637, A39

\bibitem[\protect\citeauthoryear{Bulut et al.}{2021}]{Bulut2021} 
Bulut N., Roncero O., Aguado A., Loison J.-C., Navarro-Almaida D., Wakelam V., Fuente A., et al., 2021, A\&A, 646, A5

\bibitem[\protect\citeauthoryear{Jim{\'e}nez-Escobar \& Mu{\~n}oz Caro}{2011}]{Jimenez2011} 
Jim{\'e}nez-Escobar A., Mu{\~n}oz Caro G.~M., 2011, A\&A, 536, A91

\bibitem[\protect\citeauthoryear{Palumbo, Tielens, \& Tokunaga}{1995}]{Palumbo1995} 
Palumbo M.~E., Tielens A.~G.~G.~M., Tokunaga A.~T., 1995, ApJ, 449, 674

\bibitem[\protect\citeauthoryear{Boogert et al.}{1997}]{Boogert1997} 
Boogert A.~C.~A., Schutte W.~A., Helmich F.~P., Tielens A.~G.~G.~M., Wooden D.~H., 1997, A\&A, 317, 929

\bibitem[\protect\citeauthoryear{Neufeld et al.}{2015}]{Neufeld2015} 
Neufeld D.~A., Godard B., Gerin M., Pineau des For{\^e}ts G., Bernier C., Falgarone E., Graf U.~U., et al., 2015, A\&A, 577, A49

\bibitem[\protect\citeauthoryear{Xie et al.}{2021}]{Xie2021} 
Xie J., Fuller G.~A., Li D., Chen L., Ren Z., Wu J., Duan Y., et al., 2021, SCPMA, 64, 279511

\bibitem[\protect\citeauthoryear{Goicoechea et al.}{2009}]{Goicoechea2009} 
Goicoechea J.~R., Pety J., Gerin M., Hily-Blant P., Le Bourlot J., 2009, A\&A, 498, 771

\bibitem[\protect\citeauthoryear{Kr{\v{c}}o \& Goldsmith}{2010}]{Krco2010} 
Kr{\v{c}}o M., Goldsmith P.~F., 2010, ApJ, 724, 1402

\bibitem[\protect\citeauthoryear{Loomis, et al.}{2021}]{Loomis2021} 
Loomis R.~A., Burkhardt A.~M., Shingledecker C.~N., Charnley S.~B., Cordiner M.~A., Herbst E., Kalenskii S., et al., 2021, NatAs, 5, 188

\bibitem[\protect\citeauthoryear{Agundez, et al.}{2021b}]{Agundez2021b} 
Ag{\'u}ndez M., Marcelino N., Tercero B., Cabezas C., de Vicente P., Cernicharo J., 2021, A\&A, 649, L4

\bibitem[\protect\citeauthoryear{Agundez, et al.}{2021a}]{Agundez2021a} 
Ag{\'u}ndez M., Cabezas C., Tercero B., Marcelino N., Gallego J.~D., de Vicente P., Cernicharo J., 2021, A\&A, 647, L10

\bibitem[\protect\citeauthoryear{Cernicharo et al.}{2021f}]{Cernicharo2021f} 
Cernicharo J., Ag{\'u}ndez M., Cabezas C., Marcelino N., Tercero B., Pardo J.~R., Gallego J.~D., et al., 2021, A\&A, 647, L2

\bibitem[\protect\citeauthoryear{Cernicharo et al.}{2021e}]{Cernicharo2021e} 
Cernicharo J., Cabezas C., Ag{\'u}ndez M., Tercero B., Marcelino N., Pardo J.~R., Tercero F., et al., 2021, A\&A, 647, L3

\bibitem[\protect\citeauthoryear{Cernicharo, et al.}{2021d}]{Cernicharo2021d} 
Cernicharo J., Cabezas C., Endo Y., Agundez M., Tercero B., Pardo J.~R., Marcelino N., et al., 2021, A\&A, 650, L14

\bibitem[\protect\citeauthoryear{Cernicharo, et al.}{2021c}]{Cernicharo2021c} 
Cernicharo J., Ag{\'u}ndez M., Cabezas C., Tercero B., Marcelino N., Pardo J.~R., de Vicente P., 2021, A\&A, 649, L15

\bibitem[\protect\citeauthoryear{Cernicharo, et al.}{2021b}]{Cernicharo2021b} 
Cernicharo J., Cabezas C., Ag{\'u}ndez M., Tercero B., Pardo J.~R., Marcelino N., Gallego J.~D., et al., 2021, A\&A, 648, L3

\bibitem[\protect\citeauthoryear{Cernicharo, et al.}{2021a}]{Cernicharo2021a} 
Cernicharo J., Cabezas C., Bailleux S., Margul{\`e}s L., Motiyenko R., Zou L., Endo Y., et al., 2021, A\&A, 646, L7

\bibitem[\protect\citeauthoryear{Cernicharo, et al.}{2020b}]{Cernicharo2020b} 
Cernicharo J., Marcelino N., Ag{\'u}ndez M., Endo Y., Cabezas C., Berm{\'u}dez C., Tercero B., et al., 2020, A\&A, 642, L17

\bibitem[\protect\citeauthoryear{Cernicharo, et al.}{2020a}]{Cernicharo2020a} 
Cernicharo J., Marcelino N., Pardo J.~R., Ag{\'u}ndez M., Tercero B., de Vicente P., Cabezas C., et al., 2020, A\&A, 641, L9

\bibitem[\protect\citeauthoryear{Marcelino et al.}{2021}]{Marcelino2021} 
Marcelino N., Tercero B., Ag{\'u}ndez M., Cernicharo J., 2021, A\&A, 646, L9

\bibitem[\protect\citeauthoryear{Xue, et al.}{2020}]{Xue2020} 
Xue C., Willis E.~R., Loomis R.~A., Kelvin Lee K.~L., Burkhardt A.~M., Shingledecker C.~N., Charnley S.~B., et al., 2020, ApJL, 900, L9

\bibitem[\protect\citeauthoryear{Cabezas, et al.}{2021b}]{Cabezas2021b} 
Cabezas C., Roueff E., Tercero B., Ag{\'u}ndez M., Marcelino N., de Vicente P., Cernicharo J., 2021, A\&A 650, L15

\bibitem[\protect\citeauthoryear{Cabezas et al.}{2021a}]{Cabezas2021a} 
Cabezas C., Tercero B., Ag{\'u}ndez M., Marcelino N., Pardo J.~R., de Vicente P., Cernicharo J., 2021, A\&A, 650, L9

\bibitem[\protect\citeauthoryear{Shingledecker, et al.}{2021}]{Shingledecker2021} 
Shingledecker C.~N., Lee K.~L.~K., Wandishin J.~T., Balucani N., Burkhardt A.~M., Charnley S.~B., Loomis R., et al., 2021, arXiv, arXiv:2105.03347

\bibitem[\protect\citeauthoryear{Burkhardt, et al.}{2021}]{Burkhardt2021} 
Burkhardt A.~M., Long Kelvin Lee K., Bryan Changala P., Shingledecker C.~N., Cooke I.~R., Loomis R.~A., Wei H., et al., 2021, ApJL, 913, L18

\bibitem[\protect\citeauthoryear{McGuire, et al.}{2021}]{McGuire2021} 
McGuire B.~A., Loomis R.~A., Burkhardt A.~M., Lee K.~L.~K., Shingledecker C.~N., Charnley S.~B., Cooke I.~R., et al., 2021, Sci, 371, 1265


\bibitem[\protect\citeauthoryear{Adande, et al.}{2010}]{Adande2010}
Adande G.~R., Halfen D.~T., Ziurys L.~M., Quan D., Herbst E., 2010, ApJ, 725, 561

\bibitem[\protect\citeauthoryear{Ag{\'u}ndez \& Wakelam}{2013}]{Agundez2013}
Ag{\'u}ndez M., Wakelam V., 2013, ChRv, 113, 8710

\bibitem[\protect\citeauthoryear{Aikawa, et al.}{2001}]{Aikawa2001}
Aikawa Y., Ohashi N., Inutsuka S.-. ichiro ., Herbst E., Takakuwa S., 2001, ApJ, 552, 639

\bibitem[\protect\citeauthoryear{Bron et al.}{2021}]{Bron2021} 
Bron E., Roueff E., Gerin M., Pety J., Gratier P., Le Petit F., Guzman V., et al., 2021, A\&A, 645, A28

\bibitem[\protect\citeauthoryear{Chang \& Herbst}{2012}]{Chang2012}
Chang Q., Herbst E., 2012, ApJ, 759, 147

\bibitem[\protect\citeauthoryear{Choi et al.}{2017}]{Choi2017} 
Choi Y., Lee J.-E., Bourke T.~L., Evans N.~J., 2017, ApJS, 229, 38

\bibitem[\protect\citeauthoryear{Codella et al.}{2021}]{Codella2021} 
Codella C., Bianchi E., Podio L., Mercimek S., Ceccarelli C., L{\'o}pez-Sepulcre A., Bachiller R., et al., 
2021, A\&A, 654, A52

\bibitem[\protect\citeauthoryear{Cuppen, Morata, \& Herbst}{2006}]{Cuppen2006} 
Cuppen H.~M., Morata O., Herbst E., 2006, MNRAS, 367, 1757

\bibitem[\protect\citeauthoryear{Draine}{2003}]{Draine2003} 
Draine B.~T., 2003, ARA\&A, 41, 241

\bibitem[\protect\citeauthoryear{Esplugues, et al.}{2019}]{Esplugues2019}
Esplugues G., Cazaux S., Caselli P., Hocuk S., Spaans M., 2019, MNRAS, 486, 1853

\bibitem[\protect\citeauthoryear{Elias}{1978}]{Elias1978}
Elias J.~H., 1978, ApJ, 224, 857

\bibitem[\protect\citeauthoryear{Federman, et al.}{2021}]{Federman2021} 
Federman S.~R., Rice J.~S., Ritchey A.~M., Kim H., Lacy J.~H., Goldsmith P.~F., Flagey N., et al., 2021, ApJ, 914, 59

\bibitem[\protect\citeauthoryear{Feh{\'e}r, et al.}{2016}]{Feher2016}
Feh{\'e}r O., et al., 2016, A\&A, 590, A75

\bibitem[\protect\citeauthoryear{Fuente, et al.}{2019}]{Fuente2019}
Fuente A., et al., 2019, A\&A, 624, A105

\bibitem[\protect\citeauthoryear{Garrod \& Herbst}{2006}]{Garrod2006}
Garrod R.~T., Herbst E., 2006, A\&A, 457, 927

\bibitem[\protect\citeauthoryear{Garrod, Wakelam \& Herbst}{2007}]{Garrod2007}
Garrod R.~T., Wakelam V., Herbst E., 2007, A\&A, 467, 1103

\bibitem[\protect\citeauthoryear{Ge et al.}{2020}]{Ge2020} 
Ge J., Mardones D., Inostroza N., Peng Y., 2020, MNRAS, 497, 3306

\bibitem[\protect\citeauthoryear{Gerin, et al.}{2011}]{Gerin2011}
Gerin M., Ka{\'z}mierczak M., Jastrzebska M., Falgarone E., Hily-Blant P., Godard B., de Luca M., 2011, A\&A, 525, A116

\bibitem[\protect\citeauthoryear{Gibb et al.}{2000}]{Gibb2000} 
Gibb E.~L., Whittet D.~C.~B., Schutte W.~A., Boogert A.~C.~A., Chiar J.~E., Ehrenfreund P., Gerakines P.~A., et al., 2000, ApJ, 536, 347

\bibitem[\protect\citeauthoryear{Gibb et al.}{2004}]{Gibb2004} 
Gibb E.~L., Whittet D.~C.~B., Boogert A.~C.~A., Tielens A.~G.~G.~M., 2004, ApJS, 151, 35

\bibitem[\protect\citeauthoryear{Goldsmith et al.}{2010}]{Goldsmith2010} 
Goldsmith P.~F., Velusamy T., Li D., Langer W.~D., 2010, ApJ, 715, 1370

\bibitem[\protect\citeauthoryear{Goldsmith et al.}{2021}]{Goldsmith2021} 
Goldsmith P.~F., Langer W.~D., Seo Y., Pineda J., Stutzki J., Guevara C., Aladro R., et al., 2021, ApJ, 916, 6

\bibitem[\protect\citeauthoryear{Gratier, et al.}{2016}]{Gratier2016}
Gratier P., Majumdar L., Ohishi M., Roueff E., Loison J.~C., Hickson K.~M., Wakelam V., 2016, ApJS, 225, 25

\bibitem[\protect\citeauthoryear{Herbst \& van Dishoeck}{2009}]{Herbst2009} 
Herbst E., van Dishoeck E.~F., 2009, ARA\&A, 47, 427

\bibitem[\protect\citeauthoryear{Hirahara, et al.}{1992}]{Hirahara1992}
Hirahara Y., et al., 1992, ApJ, 394, 539

\bibitem[\protect\citeauthoryear{Jenkins}{2009}]{Jenkins2009}
Jenkins E.~B., 2009, ApJ, 700, 1299

\bibitem[\protect\citeauthoryear{Kaifu et al.}{2004}]{Kaifu2004}
Kaifu N., Ohishi M., Kawaguchi K., Saito S., Yamamoto S., Miyaji T., Miyazawa K., et al., 2004, PASJ, 56, 69

\bibitem[\protect\citeauthoryear{Laas \& Caselli}{2019}]{Laas2019}
Laas J.~C., Caselli P., 2019, A\&A, 624, A108

\bibitem[\protect\citeauthoryear{Lee, et al.}{1996}]{Lee1996}
Lee H.-H., Herbst E., Pineau des Forets G., Roueff E., Le Bourlot J., 1996, A\&A, 311, 690

\bibitem[\protect\citeauthoryear{Li \& Goldsmith}{2003}]{Li2003} 
Li D., Goldsmith P.~F., 2003, ApJ, 585, 823

\bibitem[\protect\citeauthoryear{Li, et al.}{2013}]{LiXiaohu2013}
Li X., Heays A.~N., Visser R., Ubachs W., Lewis B.~R., Gibson S.~T., van Dishoeck E.~F., 2013, A\&A, 555, A14

\bibitem[\protect\citeauthoryear{Liszt, et al.}{2018}]{Liszt2018}
Liszt H., Gerin M., Beasley A., Pety J., 2018, ApJ, 856, 151

\bibitem[\protect\citeauthoryear{Loison, et al.}{2017}]{Loison2017}
Loison J.-C., et al., 2017, MNRAS, 470, 4075

\bibitem[\protect\citeauthoryear{Lucas \& Liszt}{2000}]{Lucas2000}
Lucas R., Liszt H.~S., 2000, A\&A, 358, 1069

\bibitem[\protect\citeauthoryear{Luo et al.}{2019}]{Luo2019} 
Luo G., Feng S., Li D., Qin S.-L., Peng Y., Tang N., Ren Z., et al., 2019, ApJ, 885, 82

\bibitem[\protect\citeauthoryear{Majumdar, et al.}{2017}]{Majumdar2017}
Majumdar L., et al., 2017, MNRAS, 466, 4470

\bibitem[\protect\citeauthoryear{Maffucci, et al.}{2018}]{Maffucci2018}
Maffucci D.~M., Wenger T.~V., Le Gal R., Herbst E., 2018, ApJ, 868, 41

\bibitem[\protect\citeauthoryear{McGuire, et al.}{2017}]{McGuire2017}
McGuire B.~A., Burkhardt A.~M., Shingledecker C.~N., Kalenskii S.~V., Herbst E., Remijan A.~J., McCarthy M.~C., 2017, ApJL, 843, L28

\bibitem[\protect\citeauthoryear{Navarro-Almaida et al.}{2021}]{Navarro2021} 
Navarro-Almaida D., Fuente A., Majumdar L., Wakelam V., Caselli P., Rivi{\`e}re-Marichalar P., Trevi{\~n}o-Morales S.~P., et al., 
2021, A\&A, 653, A15

\bibitem[\protect\citeauthoryear{Ohishi, Irvine \& Kaifu}{1992}]{Ohishi1992IAUS}
Ohishi M., Irvine W.~M., Kaifu N., 1992, IAUS, 150, 171, IAUS..150

\bibitem[\protect\citeauthoryear{Pineda et al.}{2010}]{Pineda2010}
Pineda J.~L., Goldsmith P.~F., Chapman N., Snell R.~L., Li D., Cambr{\'e}sy L., Brunt C., 2010, ApJ, 721, 686

\bibitem[\protect\citeauthoryear{Potapov et al.}{2020}]{Potapov2020} 
Potapov A., Bouwman J., J{\"a}ger C., Henning T., 2020, arXiv, arXiv:2008.10951

\bibitem[\protect\citeauthoryear{Poteet, Whittet, \& Draine}{2015}]{Poteet2015}
Poteet C.~A., Whittet D.~C.~B., Draine B.~T., 2015, ApJ, 801, 110

\bibitem[\protect\citeauthoryear{Przybilla, Nieva, \& Butler}{2008}]{Przybilla2018} 
Przybilla N., Nieva M.-F., Butler K., 2008, ApJL, 688, L103

\bibitem[\protect\citeauthoryear{Ruaud, et al.}{2015}]{Ruaud2015}
Ruaud M., Loison J.~C., Hickson K.~M., Gratier P., Hersant F., Wakelam V., 2015, MNRAS, 447, 4004

\bibitem[\protect\citeauthoryear{Ruaud, Wakelam \& Hersant}{2016}]{Ruaud2016}
Ruaud M., Wakelam V., Hersant F., 2016, MNRAS, 459, 3756

\bibitem[\protect\citeauthoryear{Sakai, et al.}{2012}]{Sakai2012}
Sakai N., Maezawa H., Sakai T., Menten K.~M., Yamamoto S., 2012, A\&A, 546, A103

\bibitem[\protect\citeauthoryear{Semenov, et al.}{2010}]{Semenov2010}
Semenov D., et al., 2010, A\&A, 522, A42

\bibitem[\protect\citeauthoryear{Seo, et al.}{2019}]{Seo2019}
Seo Y.~M., et al., 2019, ApJ, 871, 134

\bibitem[\protect\citeauthoryear{Snow \& McCall}{2006}]{Snow2006}
Snow T.~P., McCall B.~J., 2006, ARA\&A, 44, 367

\bibitem[\protect\citeauthoryear{Soma, et al.}{2015}]{Soma2015}
Soma T., Sakai N., Watanabe Y., Yamamoto S., 2015, ApJ, 802, 74

\bibitem[\protect\citeauthoryear{Soma, et al.}{2018}]{Soma2018}
Soma T., Sakai N., Watanabe Y., Yamamoto S., 2018, ApJ, 854, 116

\bibitem[\protect\citeauthoryear{Suutarinen, et al.}{2011}]{Suutarinen2011}
Suutarinen A., et al., 2011, A\&A, 531, A121

\bibitem[\protect\citeauthoryear{Suzuki, et al.}{1992}]{Suzuki1992}
Suzuki H., Yamamoto S., Ohishi M., Kaifu N., Ishikawa S.-I., Hirahara Y., Takano S., 1992, ApJ, 392, 551

\bibitem[\protect\citeauthoryear{Thiel et al.}{2017}]{Thiel2017}
Thiel V., Belloche A., Menten K.~M., Garrod R.~T., M{\"u}ller H.~S.~P., 2017, A\&A, 605, L6

\bibitem[\protect\citeauthoryear{Thiel, et al.}{2019}]{Thiel2019}
Thiel V., et al., 2019, A\&A, 623, A68

\bibitem[\protect\citeauthoryear{Turner}{1998}]{Turner1998} 
Turner B.~E., 1998, ApJ, 501, 731

\bibitem[\protect\citeauthoryear{Turner, Terzieva, \& Herbst}{1999}]{Turner1999} 
Turner B.~E., Terzieva R., Herbst E., 1999, ApJ, 518, 699

\bibitem[\protect\citeauthoryear{Turner}{2000}]{Turner2000}
Turner B.~E., 2000, ApJ, 542, 837

\bibitem[\protect\citeauthoryear{Vastel, et al.}{2018}]{Vastel2018}
Vastel C., et al., 2018, MNRAS, 478, 5514

\bibitem[\protect\citeauthoryear{Vidal, et al.}{2017}]{Vidal2017}
Vidal T.~H.~G., Loison J.-C., Jaziri A.~Y., Ruaud M., Gratier P., Wakelam V., 2017, MNRAS, 469, 435

\bibitem[\protect\citeauthoryear{Visser, van Dishoeck \& Black}{2009}]{Visser2009}
Visser R., van Dishoeck E.~F., Black J.~H., 2009, A\&A, 503, 323

\bibitem[\protect\citeauthoryear{Wakelam \& Herbst}{2008}]{Wakelam2008} 
Wakelam V., Herbst E., 2008, ApJ, 680, 371

\bibitem[\protect\citeauthoryear{Wakelam, et al.}{2015}]{Wakelam2015}
Wakelam V., et al., 2015, ApJS, 217, 20

\bibitem[\protect\citeauthoryear{Womack, Ziurys, \& Wyckoff}{1992}]{Womack1992} 
Womack M., Ziurys L.~M., Wyckoff S., 1992, ApJ, 393, 188

\bibitem[\protect\citeauthoryear{Xu, et al.}{2016a}]{Xu2016a}
Xu D., Li D., Yue N., Goldsmith P.~F., 2016, ApJ, 819, 22

\bibitem[\protect\citeauthoryear{Xu \& Li}{2016b}]{Xu2016b}
Xu D., Li D., 2016, ApJ, 833, 90

\bibitem[\protect\citeauthoryear{Zuo et al.}{2018}]{Zuo2018} 
Zuo P., Li D., Peek J.~E.~G., Chang Q., Zhang X., Chapman N., Goldsmith P.~F., et al., 2018, ApJ, 867, 13



\end{thebibliography}



\end{document}